\documentclass[english]{article}
\usepackage[T1]{fontenc}
\usepackage[latin9]{inputenc}
\usepackage{geometry}
\usepackage{color}
\usepackage{array}
\usepackage{float}
\usepackage{amsthm,amssymb}
\usepackage{amsmath}
\usepackage{graphicx}
\usepackage[labelsep=period]{caption}

\date{}
\makeatletter

\newcommand{\lyxmathsym}[1]{\ifmmode\begingroup\def\b@ld{bold}
  \text{\ifx\math@version\b@ld\bfseries\fi#1}\endgroup\else#1\fi}

\providecommand{\tabularnewline}{\\}
\floatstyle{ruled}
\newfloat{algorithm}{tbp}{loa}
\providecommand{\algorithmname}{Algorithm}
\floatname{algorithm}{\protect\algorithmname}

\numberwithin{equation}{section}
\numberwithin{figure}{section}
\theoremstyle{plain}
\newtheorem{thm}{\protect\theoremname}
  \theoremstyle{definition}
  \newtheorem{defn}[thm]{\protect\definitionname}

\makeatother

\usepackage{babel}
  \providecommand{\definitionname}{Definition}
\providecommand{\theoremname}{Theorem}

\newcommand{\beginsupplement}{%
        \setcounter{table}{0}
        \renewcommand{\thetable}{S\arabic{table}}%
        \setcounter{figure}{0}
	\renewcommand{\figurename}{Fig.}
        \renewcommand{\thefigure}{S\arabic{figure}}%
     }
\renewcommand{\figurename}{Image}

\begin{document}
\title{A Rewritable, Random-Access DNA-Based Storage System}

\author{S. M. Hossein Tabatabaei Yazdi$^{1\dagger}$, Yongbo Yuan$^{2\dagger}$, Jian
Ma$^{3,4}$, Huimin Zhao$^{2,4}$,\\ Olgica Milenkovic$^{1*}$}

\maketitle

\section*{Affiliations: }
\begin{description}
  \item $^{1}$Department of Electrical and Computer Engineering, University
of Illinois, Urbana, IL 61801
  \item $^{2}$Department of Chemical and Biomolecular Engineering, University
of Illinois, Urbana, IL 61801 
  \item $^{3}$Department of Bioengineering, University of Illinois, Urbana,
IL 61801
\item $^{4}$Institute for Genomic Biology, University of Illinois, Urbana,
IL 61801
\item $^{\dagger}$These authors contributed equally to the work.
\item $^{*}$To whom correspondences should be addressed: Olgica Milenkovic, e-mail:
milenkov@illinois.edu   
\end{description}

\author{\sloppy}
\begin{abstract}
We describe the first DNA-based storage architecture that enables
random access to data blocks and rewriting of information stored at
arbitrary locations within the blocks. The newly developed architecture
overcomes drawbacks of existing read-only methods that require decoding
the whole file in order to read one data fragment. Our system is based
on new constrained coding techniques and accompanying DNA editing methods that 
ensure data reliability, specificity and sensitivity of access, and at the same time provide 
exceptionally high data storage capacity.
As a proof of concept, we encoded parts of the Wikipedia pages of six 
universities in the USA, and selected and edited parts of the text written in DNA corresponding
to three of these schools. The results suggest that DNA is a versatile media suitable for both ultrahigh density archival and rewritable storage applications. 
\end{abstract}
\maketitle

Addressing the emerging demands for massive data repositories, and building upon the rapid development of technologies for DNA synthesis
and sequencing, a number of laboratories have recently outlined architectures for archival DNA-based storage~\cite{bancroft2001long,davis1996microvenus,church2012next,goldman2013towards,grass2015robust}.
The architecture in~\cite{church2012next} achieved a storage density
of $700$ TB/gram, while the system described in~\cite{goldman2013towards}
raised the density to $2.2$ PB/gram. The success of the latter method may be largely attributed to three 
\emph{classical coding schemes:} Huffman
coding, differential coding, and single parity-check coding~\cite{goldman2013towards}.
Huffman coding was used for data compression, while differential coding was used for eliminating homopolymers (i.e., repeated consecutive bases) in the DNA strings. Parity-checks were used to add 
controlled redundancy, which in conjunction with four-fold coverage allows for mitigating assembly errors\footnote{Another class of DNA error-correcting schemes based on Reed-Solomon (RS) codes was recently reported in~\cite{grass2015robust}.}.

Due to dynamic changes in biotechnological systems, none of the three coding schemes represents a suitable solution from the perspective of current DNA sequencer designs: Huffman codes are fixed-to-variable
length compressors that can lead to catastrophic error propagation
in the presence of sequencing noise; the same is true of differential
codes. Homopolymers do not represent a significant source
of errors in Illumina sequencing platforms~\cite{ross2013characterizing}, while 
single parity redundancy or RS codes and differential encoding are inadequate for combating error-inducing sequence patterns such as long substrings with high GC content~\cite{ross2013characterizing}. 
As a result, assembly errors are likely, and were observed
during the readout process described in~\cite{goldman2013towards}. 

An even more important issue that prohibits the practical wide-spread
use of the schemes described in~\cite{church2012next,goldman2013towards} is that accurate partial and random access to data is impossible, as one has to reconstruct the whole text in order to read or retrieve the information encoded even in a few bases. Furthermore, all current designs support read-only storage. The first limitation represents a significant drawback, as one usually needs to accommodate access to specific data sections;
the second limitation prevents the use of current DNA storage methods in architectures that call for moderate data editing, for storing frequently updated information and memorizing the history of edits.
Moving from a read-only to a rewritable DNA storage system requires a major implementation paradigm shift, as: 

\textbf{1.} Editing in the compressive domain may require rewriting almost the whole information content; 

\textbf{2.} Rewriting is complicated by the current data DNA storage format that involves reads of length $100$ bps shifted by $25$ bps so as to ensure four-fold coverage of the sequence
(See Figure~\ref{fig:system} (a) for an illustration and description of the data format used in~\cite{goldman2013towards}). 
In order to rewrite one base, one needs to selectively access and modify four ``consecutive'' reads; 

\textbf{3.} Addressing methods used in~\cite{church2012next,goldman2013towards} only allow for determining the position of a read in a file, but 
cannot ensure precise selection of reads of interest, as undesired cross-hybridization between the primers and parts of the information blocks may occur.

To overcome the aforementioned issues, we developed a new, random-access
and rewritable DNA-based storage architecture based on DNA sequences endowed with specialized address strings that may be used for selective information access and encoding with inherent error-correction capabilities. The addresses are designed to be \emph{mutually uncorrelated} and to satisfy the \emph{error-control running digital sum constraint}~\cite{cohen1991dc,blaum1993error}. Given the address sequences, encoding is performed by stringing together properly terminated prefixes of the addresses as dictated by the information sequence. This encoding method represents a special form of \emph{prefix-synchronized coding}~\cite{gilbert1960synchronization}.
Given that the addresses are chosen to be uncorrelated and at large Hamming distance from each other, 
it is highly unlikely for one address to be confused with another address or with another section of the encoded blocks. Furthermore, 
selection of the blocks to be rewritten is made possible by the prefix encoding format, while rewriting is performed via two DNA editing techniques, the gBlock and OE-PCR (overlap-extension polymerase chain reaction) methods~\cite{packergblocks,bryksin2010overlap}. 
With the latter method, rewriting is done in several steps by using short and cheap primers. The first method is more efficient, but requires synthesizing longer and hence more expensive primers. Both methods were tested on DNA encoded Wikipedia entries of size $17$ KB, corresponding to six universities, where information in one, two and three blocks was rewritten in the DNA encoded domain. The rewritten blocks were selected, amplified and Sanger sequenced~\cite{schuster2008next} to verify that selection and rewriting are performed with $100\%$ accuracy.

\section{Results} \label{sec:results}

The main feature of our storage architecture that enables highly sensitive random access and accurate rewriting is \emph{addressing}.
The rational behind the proposed approach is that each block in a random access system must be equipped with an address that will allow for unique selection and amplification via DNA sequence primers.

\begin{figure*}[ht]
\begin{center}
\centerline{\includegraphics[width=0.85\textwidth]{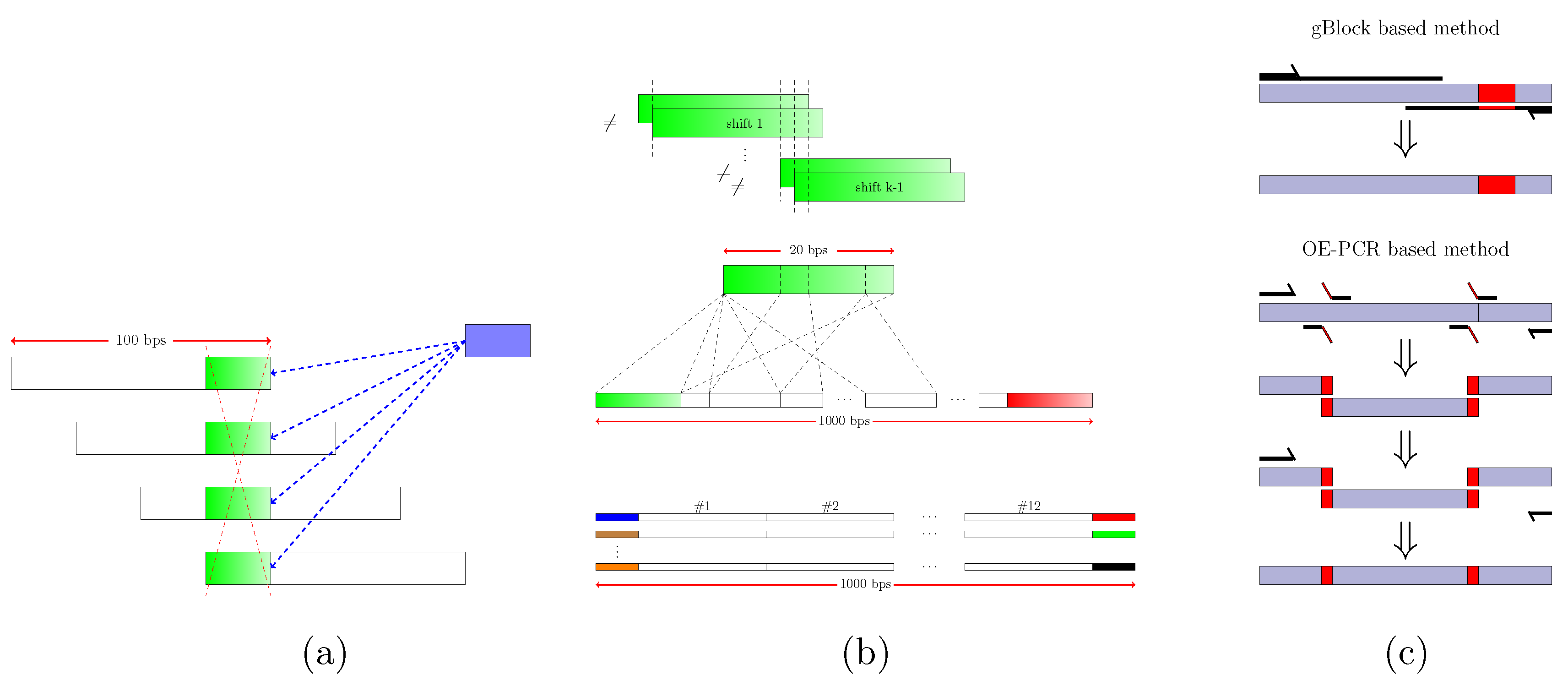}}
\caption{(a) The scheme of \cite{goldman2013towards} uses a
storage format consisting of DNA strings that cover the encoded compressed
text in fragments of length of $100$ bps. The fragments overlap in
$75$ bps, thereby providing $4$-fold coverage for all except the
flanking end bases. This particular fragmenting procedure prevents efficient 
file editing: If one were to rewrite the \textquotedblleft{}shaded\textquotedblright{}
block, all four fragments containing this block would need to be selected
and rewritten at different positions to record the new \textquotedblleft{}shaded\textquotedblright{}
block. (b) The address sequence construction process using the notions of \emph{autocorrelation
and cross-correlation of sequences}~\cite{immink2004codes}. A sequence
is uncorrelated with itself if no proper prefix of the sequence is
also a suffix of the same sequence. Alternatively, no shift of the sequence overlaps with the sequence itself. 
Similarly, two different sequences are uncorrelated if no prefix of one sequence matches a suffix of the
other. Addresses are chosen to be mutually uncorrelated, and each
$1000$ bps block is flanked by an address of length $20$ on the left and by another
address of length $20$ on the right (colored ends). (c) Content rewriting via DNA editing: the gBlock
method~\cite{packergblocks} for short rewrites, and the cost efficient OE-PCR (Overlap
Extension-PCR) method~\cite{bryksin2010overlap} for sequential rewriting of longer blocks.}\label{fig:system}
\end{center}
\end{figure*}


Instead of storing blocks mimicking the structure and length of reads generated
during high-throughput sequencing, we synthesized blocks of length $1000$ bps tagged at both ends
by specially designed address sequences. Adding addresses to short blocks of length $100$ bps would incur a large storage overhead, while synthesizing blocks longer than $1000$ bps using current technologies is prohibitively costly. 

More precisely, each data block of length $1000$ bps was flanked at both ends by two unique, yet different, address blocks
of length $20$ bps. These addresses are used to provide specificity of access (see Figure~\ref{fig:system} (b) and the Supplementary Information for details). The remaining $960$ bases in a block are divided into
$12$ sub-blocks of length $80$ bps, with each block encoding six
words of the text. The ``word-encoding'' process may be seen as a specialized compaction scheme suitable for rewriting, and it operates as follows. First, different words in the text are counted and tabulated in a dictionary. Each word in the dictionary is converted into a binary sequence of length sufficiently long to allow for encoding of the dictionary. For our current implementation and texts of choice, described in the Supplementary Information section, this length was set to $24$. Encodings of six consecutive words are subsequently grouped into binary sequences of length $144$. The two-bit $11$ is appended as a word marker to the left hand side of each binary sequence of length $144$, resulting in sequences of length $146$ bits. The binary sequences are subsequently translated into DNA blocks of length $80$ bps using a new family of DNA prefix-synchronized codes described in the Methods section. Our choice for the number of jointly encoded words is governed by the goal to make rewrites as straightforward as possible and to avoid error propagation due to variable codelengths. Furthermore, as most rewrites include words, rather than individual symbols, the word encoding method represents an efficient means for content update. Details regarding the counting and grouping procedure may be found in the Supplementary Information.

For three selected access queries, the $1000$ bps blocks containing the
desired information were identified via primers corresponding to their
unique addresses, PCR amplified, Sanger sequenced, and subsequently decoded.

Two methods were used for content rewriting. 
If the region to be rewritten had length exceeding several hundreds, new sequences with unique primers 
were synthesized as this solution represents a less costly
alternative to rewriting. For the case that a relatively short substring of the
encoded string had to be modified, the corresponding $1000$ bps block hosting the string was identified and the 
changes were generated via DNA editing.

Both the random access and rewriting protocols were tested experimentally
on two jointly stored text files. One text file, of size $4$ KB, contained the history
of University of Illinois, Urbana-Champaign (UIUC) based on its Wikipedia
entry retrieved on $12/15/2013$. The other text file, of size $13$ KB,
contained the introductory Wikipedia entries of Berkeley, Harvard, MIT,
Princeton, and Stanford, retrieved on $04/27/2014$. 

\begin{figure*}[ht]
\begin{center}
\centerline{\includegraphics[width=0.85\textwidth]{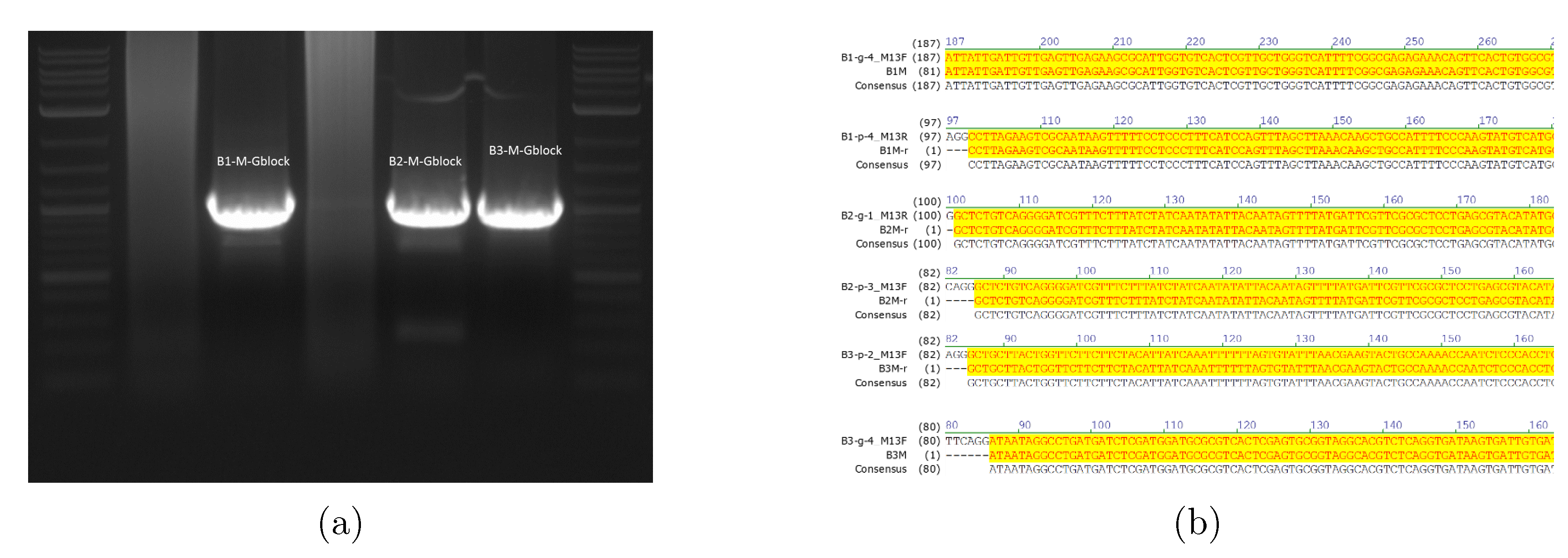}}
\caption{(a) Gel electrophoresis results for three blocks, indicating that the length of the three selected and amplified sequences is tightly concentrated around $1000$ bps. (b) Output of the Sanger sequencer, where all bases shaded in yellow correspond to correct readouts. The sequencing results confirmed that the desired sequences were selected, amplified, and rewritten with 100$\%$ accuracy.}\label{fig:electro}
\end{center}
\end{figure*}

Encoded information was converted into DNA blocks of length $1000$
bps synthesized by IDT (Integrated DNA Technologies), at a cost of
$\$149$ per $1000$ bps (see http://www.idtdna.com/pages/products/genes/gblocks-gene-fragments).
The rewriting experiments encompassed:

\textbf{1.} \emph{PCR selection and amplification of one $1000$ bps sequence and simultaneous selection and amplification of three $1000$ bps sequences in
the pool.} All $32$ linear $1000$ bps fragments were mixed, and the mixture
was used as a template for PCR amplification and selection. The results of amplification were verified by confirming
sequence lengths of $1000$ bps banks via gel electrophoresis (Figure~\ref{fig:electro} (a)) and 
by randomly sampling $3$-$5$ sequences from the pools and Sanger sequencing them (Figure~\ref{fig:electro} (b)).

\textbf{2.} \emph{Experimental content rewriting via synthesis of edits located
at various positions in the $1000$ bps blocks.} For simplicity of notation, we refer to the blocks in the pool on which we performed selection and editing as B1, B2, and B3. Two primers were synthesized for each rewrite in the blocks, for the forward and reverse direction. In addition, two different editing/mutation techniques were used, gBlock and Overlap-Extension (OE) PCR. gBlocks are double-stranded genomic fragments used as primers or for the purpose of genome editing, while OE-PCR is a variant of PCR used for specific DNA sequence editing via point editing/mutations or splicing. To demonstrate the plausibility of a cost efficient method for editing, OE-PCR was implemented with general primers ($\leq60$ bps) only. Note that for edits shorter than $40$ bps, the mutation sequences were designed as overhangs in primers.
Then, the three PCR products were used as templates for the final PCR reaction involving the entire $1000$ bps rewrite. Figure \ref{fig:system} (c) illustrates the described rewriting process. 
In addition, a summary of the experiments performed is provided in Table~\ref{tab:summary}.

\begin{table}
\centering{}%
 \begin{tabular}{|| c | c | c | c || } \hline
Sequence identifier - Editing Method & $\#$ of sequence samples & Length of edits (bps) & Selection accuracy/error percentage\\
\hline 
B1-M-gBlock & $5$ & $20$ & $(5/5) /0\%$ \\
\hline 
B1-M-PCR & $5$ & $20$ & $(5/5) /0\%$ \\
\hline 
B2-M-gBlock & $5$ & $28$ & $(5/5) /0\%$  \\
\hline 
B2-M-PCR & $5$ & $28$ & $(5/5) /0\%$  \\
\hline 
B3-M-gBlock & $5$ & $41+29$ & $(5/5) /0\%$ \\
\hline 
B3-M-PCR & $5$ & $41+29$ & $(5/5) /0\%$ \\
\hline 
\end{tabular}
\vspace{0.1in}
\caption{\label{tab:summary} Selection, rewriting and sequencing results. Each rewritten
$1000$ bps sequence was ligated to a linearized pCRTM-Blunt vector
using the Zero Blunt PCR Cloning Kit and was transformed into \emph{E.
coli.} The \emph{E. coli} strains with correct plasmids were sequenced
at ACGT, Inc. Sequencing was performed using two universal primers:
M13F\_20 (in the reverse direction) and M13R (in the forward direction)
to ensure that the entire block of $1000$ bps is covered.}
\end{table}

Given that each nucleotide has weight roughly equal to $650$ daltons ($650\times1.67\times10^{-24}$
grams), and given that $27,000+5000=32,000$ bps were needed to encode a file
of size $13+4=17$ KB in ASCII format, we estimate a potential storage density of 
$4.9\times10^{20}$ B/g. This density significantly surpasses the current state-of-the-art storage density of $2.2\times10^{15}$
bytes/g, as we avoid costly multiple coverage, use larger blocklengths and specialized word encoding schemes. A performance comparison of the three currently known DNA-based storage media is given in Table~\ref{tab:2}. We observe 
that the cost of sequence synthesis in our storage model
is significantly higher than the corresponding cost of the prototype
in~\cite{goldman2013towards}, as blocks of length $1000$ bps are
still difficult to synthesize. This trend it likely to change dramatically
in the near future, as within the last seven months, the cost of synthesizing
$1000$ bps blocks reduced almost $7$-fold. Despite its high cost, our system
offers exceptionally large storage density, and for the first time,
enables random access and content rewriting features. Furthermore,
although we used Sanger sequencing methods for our small scale experiment, for large scale
storage projects Next Generation Sequencing (NGS) technologies will enable significant reductions in readout costs.

\begin{table}
\centering{}%
  \begin{tabular}{|| c | c | c | c || } \hline
                                         & Church et.al. \cite{church2012next} & Goldman et.al. \cite{goldman2013towards} & Our scheme \\ \hline
    Density  & $0.7 \times 10^{15}$ B/g & $2.2 \times 10^{15}$ B/g & $4.9 \times 10^{20}$ B/g  \\  \hline
    File size & $5.27$MB & $739$KB & File size: $17$KB \\   \hline
    Cost       & Not available & $\$12,600$ & $\$4,023$ \\  \hline 
    Features       & Archival, no random-access & Archival, no random-access & Rewritable, random-access \\  \hline 
\end{tabular}
\vspace{0.12in}
\caption{\label{tab:2} Comparison of storage densities for the DNA \emph{encoded} information
expressed in B/g (bytes per gram), file size, synthesis cost, and random
access features of three known DNA storage technologies. Note that
the density does not reflect the entropy of the information
source, as the text files are encoded in ASCII format, which is a
redundant representation system.}
\end{table}


\section{Methods} \label{sec:methods}


\subsection{Address Design and Encoding} \label{sec:coding}

To encode information on DNA media, we employed a two-step procedure. First, we designed address sequences of short length which satisfy a number of constraints that makes them suitable for highly selective random access~\cite{immink2004codes}. \emph{Constrained coding} ensures that DNA patterns prone to sequencing errors are avoided and that DNA blocks are accurately accessed, amplified and selected without perturbing or accidentally selecting other blocks in the DNA pool. The coding constraints apply to address primer design, but also indirectly govern the properties of the fully encoded DNA information blocks. The design procedure used is semi-analytical, in so far that it combines combinatorial methods with computer search techniques.

We required the address sequences to satisfy the following constraints:
\begin{itemize}
\item (C1) Constant GC content (close to $50\%$) of all their prefixes
of sufficiently long length. DNA strands with
$50\%$ GC content are more stable than DNA strands with lower or
higher GC content and have better coverage during sequencing. Since encoding user information is accomplished
via prefix-synchronization, it is important to impose the GC content constraint on the addresses as well as their prefixes, as the latter requirement also ensures that all fragments of encoded data blocks have balanced GC content.
\item (C2) Large mutual Hamming distance, as it reduces the probability
of erroneous address selection. Recall that the Hamming distance between
two strings of equal length equals the number of positions at which the
corresponding symbols disagree. 
An appropriate choice for the minimum Hamming distance is equal to half of the address sequence length ($10$ bps in our current implementation which uses length $20$ address primers).
\item (C3) Uncorrelatedness of the addresses, which imposes the restriction that prefixes of one address do not appear as suffixes of the same or another address and vice versa. The motivation for this new constraint comes from the fact that addresses are used to provide unique identities for the blocks, and that their substrings should therefore not appear in ``similar form'' within other addresses. Here, ``similarity'' is assessed in terms of hybridization affinity. Furthermore, long undesired prefix-suffix matches may lead to read assembly errors in blocks during joint informational retrieval and sequencing.
\item (C4) Absence of secondary (folding) structures, as such structures may cause errors
in the process of PCR amplification and fragment rewriting. 
\end{itemize}


Addresses satisfying constraints C1-C2 may be constructed via error-correcting codes with small running digital
sum~\cite{cohen1991dc} adapted for the new storage system. Properties of these codes are discussed in Section~\ref{sub:balanced}. The novel notion of \emph{mutually uncorrelated sequences} is introduced in~\ref{sub:correlation}. 
Constructing addresses that simultaneously satisfy the constraints C1-C4 and determining bounds on the largest number of such sequences is prohibitively complex~\cite{morita1996construction,milenkovic2006design}. To mitigate this problem, we resort to a \emph{semi-constructive} address design approach, in which balanced error-correcting codes are designed independently, and subsequently expurgated so as to identify a large set of mutually uncorrelated sequences. The resulting sequences are subsequently tested for secondary structure using~\emph{mfold} and~\emph{Vienna}~\cite{rouillard2003oligoarray}. We conjecture that the number of sequences satisfying C1-C4 grows exponentially with their length: proofs towards establishing this claim include results on the exponential size of codes under each constraint individually.

Given two uncorrelated sequences as flanking addresses of one block, one of the sequences is selected to encode
user information via a new implementation of \emph{prefix-synchronized encoding}~\cite{guibas1978maximal,rouillard2003oligoarray}, described in~\ref{sub:prefix}.
The asymptotic rate of an optimal single sequence prefix-free codes is one. Hence, there is no asymptotic coding loss for avoiding prefixes of one sequence; we only observe a minor coding loss for each finite-length block. For multiple sequences of arbitrary structure, the problem of determining the optimal code rate is significantly more complicated and the rates have to be evaluated numerically, by solving systems of linear equations~\cite{guibas1978maximal} as described in~\ref{sub:prefix} and the Supplementary Information. This system of equations leads to a particularly simple form for the generating function of mutually uncorrelated sequences, as explained in the Supplementary Information. 

\subsection{Balanced Codes and Running Digital Sums} \label{sub:balanced} 

An important criteria for selecting block addresses is to ensure
that the corresponding DNA primer sequences have prefixes with a GC content approximately
equal to $50\%$, and that the sequences are at large pairwise Hamming distance. 
Due to their applications in optical storage, codes that address related issues have been studied in 
a different form under the name of \emph{bounded running digital sum} (BRDS) codes~\cite{cohen1991dc,blaum1993error}. A detailed overview of this coding technique may be found in~\cite{cohen1991dc}.

Consider a sequence $a=a_{0},a_{1},a_{2},\ldots,a_{l},\ldots,a_n$ over the alphabet $\{{-1,1\}}$. 
We refer to $S_{l}\left(a\right)=\sum_{i=0}^{l-1}a_{i}$ as the running digital sum (RDS) of the sequence $a$ up to length $l$, $l\geq0$. Let
$D_{a}=\max\left\{ \left|S_{l}\left(a\right)\right|:l\geq0\right\}$ denote the largest value of the running digital sum of the sequence $a$. 
For some predetermined value $D>0$, a set of sequences $\{{a(i)\}}_{i=1}^{M}$ is termed a BRDS code with parameter $D$ if $D_{a(i)}\leq D$ for all $i=1,\ldots,M$. Note that one can define non-binary BRDS codes in an equivalent manner, with the alphabet usually assumed to be symmetric, $\{{-q,-q+1,\ldots,-1,1,\ldots, q-1,q\}}$, and where $q \geq 1$. A set of DNA sequences over $\left\{ \mathtt{A,T,G,C}\right\}$ may be constructed in a straightforward manner by mapping each $+1$ symbol into one of the bases $\left\{ \mathtt{A,T}\right\},$ and $-1$ into one of the bases $\left\{ \mathtt{G,C}\right\}$, or vice versa. Alternatively, one can use BRDS over an alphabet of size four directly.

To address the constraints C1-C2, one needs to construct a large set of BRDS codewords at sufficiently large Hamming distance from each other. Via the mapping described above, these codewords may be subsequently translated to DNA sequences with a GC content approximately equal to $50\%$ for all sequence prefixes, and at the same Hamming distance as the original sequences.

Let $\left(n,C,d;D\right)$ be the parameters of a BRDS error-correcting
code, where $C$ denotes the number of codewords of length $n$, $d$ denotes the minimum distance of the code, while $\frac{\log C}{n}$ equals the code rate. 
For $D=1$ and $d=2$, the best known BRDS-code has parameters
$\left(n,2^{\frac{n}{2}},2;1\right)$, while for $D=2$ and $d=1$, codes with parameters $\left(n,3^{\frac{n}{2}},1;2\right)$ exist.
For $D=2$ and $d=2$, the best known BRDS code has
parameters $\left(n,2\cdot3^{\left(\frac{n}{2}\right)-1},2;2\right)$~\cite{blaum1993error}.
Note that each of these codes has an exponentially large number of
codewords, among which a (sufficiently) large number of sequences satisfy the required correlation property C3, discussed next, and the folding
property C4. Codewords satisfying constraints C3-C4 were found by expurgating the BRDS codes via computer search.

\subsection{Sequence Correlation} \label{sub:correlation}

We describe next the notion of autocorrelation of a sequence 
and introduce the related notion of mutual correlation of sequences. 

It was shown in \cite{guibas1978maximal} that the
autocorrelation function is the crucial mathematical concept for studying
sequences avoiding forbidden strings and substrings. In the storage context, forbidden strings correspond to the addresses of 
the blocks in the pool. In order to accommodate
the need for selective retrieval of a DNA block without accidentally selecting any undesirable blocks, we find it necessary 
to also introduce the notion of mutually uncorrelated sequences. 

Let $X$ and $Y$ be two words, possibly of different lengths, over
some alphabet of size $q>1$. The correlation of $X$ and $Y$, denoted
by $X\circ Y$, is a binary string of the same length as $X$. The
$i$-th bit (from the left) of $X\circ Y$ is determined by placing
$Y$ under $X$ so that the leftmost character of $Y$ is under the
$i$-th character (from the left) of $X$, and checking whether the characters
in the overlapping segments of $X$ and $Y$ are identical. If they
are identical, the $i$-th bit of $X\circ Y$ is set to $1$, otherwise,
it is set to $0$. For example, for $X=\mathtt{CATCATC}$ and $Y=\mathtt{ATCATCGG}$, $X\circ Y=0100100$, as depicted below.

Note that in general, $X\circ Y\neq Y\circ X$, and that the two correlation
vectors may be of different lengths. In the example above, we have
$Y\circ X=00000000$. The autocorrelation of a word $X$ equals $X\circ X$.

In the example below, $X\circ X=1001001$. 

\[
\begin{array}{cccccccccccccccc}
X= & \mathtt{C} & \mathtt{A} & \mathtt{T} & \mathtt{C} & \mathtt{A} & \mathtt{T} & \mathtt{C}\\
Y= & \mathtt{A} & \mathtt{T} & \mathtt{C} & \mathtt{A} & \mathtt{T} & \mathtt{C} & \mathtt{G} & \mathtt{G} &  &  &  &  &  &  & 0\\
 &  & \mathtt{A} & \mathtt{T} & \mathtt{C} & \mathtt{A} & \mathtt{T} & \mathtt{C} & \mathtt{G} & \mathtt{G} &  &  &  &  &  & 1\\
 &  &  & \mathtt{A} & \mathtt{T} & \mathtt{C} & \mathtt{A} & \mathtt{T} & \mathtt{C} & \mathtt{G} & \mathtt{G} &  &  &  &  & 0\\
 &  &  &  & \mathtt{A} & \mathtt{T} & \mathtt{C} & \mathtt{A} & \mathtt{T} & \mathtt{C} & \mathtt{G} & \mathtt{G} &  &  &  & 0\\
 &  &  &  &  & \mathtt{A} & \mathtt{T} & \mathtt{C} & \mathtt{A} & \mathtt{T} & \mathtt{C} & \mathtt{G} & \mathtt{G} &  &  & 1\\
 &  &  &  &  &  & \mathtt{A} & \mathtt{T} & \mathtt{C} & \mathtt{A} & \mathtt{T} & \mathtt{C} & \mathtt{G} & \mathtt{G} &  & 0\\
 &  &  &  &  &  &  & \mathtt{A} & \mathtt{T} & \mathtt{C} & \mathtt{A} & \mathtt{T} & \mathtt{C} & \mathtt{G} & \mathtt{G} & 0
\end{array}
\]

\begin{defn}
\emph{A sequence X is self-uncorrelated if} $X\circ X=10\ldots0$. A set of sequences $\{{X_1,X_2,\ldots,X_m\}}$ is
termed mutually uncorrelated if each sequence is self-uncorrelated and if all pairs of distinct sequences satisfy 
$X_i\circ X_j=0\ldots0$ and $X_j\circ X_i=0\ldots0$.
\end{defn}
Intuitively, correlation captures the extent to which prefixes of sequences overlap with suffixes of the same or other sequences. Furthermore, the notion of mutual uncorrelatedness may be relaxed by requiring that 
only sufficiently long prefixes do not match sufficiently long suffixes of other sequences. Sequences with this property, and at sufficiently large Hamming distance, eliminate undesired address cross-hybridization during selection and cross-sequence assembly errors.

We proved the following bound on the size of the largest mutually uncorrelated set of sequences of length $n$ over an alphabet of size $q=4$. The bounds show that there exist exponentially many mutually uncorrelated sequences for any choice of $n$, and the lower bound is constructive. Furthermore, the construction used in the bound ``preserves'' the Hamming distance (see the Supplementary Information).

\begin{thm}\label{thm:bounds}
Suppose that $\left\{ X_{1},\ldots,X_{m}\right\} $ is a set of $m$ pairwise mutually
uncorrelated sequences of length $n$. Let $u\left(n\right)$ denote the largest possible value of $m$ for a given $n$.
Then
\label{lem:upper_bound}
\[
4\cdot3^{\frac{n}{4}} \leq u\left(n\right)\leq9\cdot4^{n-2}.
\]
\end{thm}
As an illustration, for $n=20$, the \emph{lower bound} equals $972$.
The proof of the theorem is give in the Supplementary Information.

It remains an open problem to determine the largest number of address sequences that jointly satisfy the constraints C1-C4. 
We conjecture that the number of such sequences is exponential in $n$, as the numbers of words that satisfy C1-C2, C3 and C4~\cite{milenkovic2006design} are exponential. Exponentially large families of address sequences are important indicators of the scalability of the system and they also influence the rate of information encoding in DNA.

Using a casting of the address sequence design problem in terms of a simple and efficient greedy search procedure, we were able to identify $1149$ sequences for length $n=20$ that satisfy constraints C1-C4, out of which $32$ pairs were used for block addressing. Another means to generate large sets of sequences satisfying the constraints is via approximate solvers for the \emph{largest independent set problem}~\cite{berman1994approximating}. Examples of sequences constructed in the aforementioned manner and used in our experiments are listed in the Supplementary Information. 

\subsection{Prefix-Synchronized DNA Codes} \label{sub:prefix}

In the previous sections, we described how to construct address sequences that can serve as unique identifiers of the blocks they are associated with. We also pointed out that once such address sequences are identified, user information has to be encoded in order to \emph{avoid} the appearance of any of the addresses, sufficiently long substrings 
of the addresses, or substrings similar to the addresses in the resulting DNA codeword blocks. For this purpose, we developed new prefix-synchronized encoding schemes based on~\cite{morita1996construction}.

To address the problem at hand, we start by introducing comma free and prefix-synchronized codes which allow for constructing codewords that avoid address patterns. A block code $\mathcal{C}$ comprising a set of codewords of length $N$ over an alphabet of size $q$ is called \emph{comma free} if and only if for any pair of not necessarily distinct codewords $a_{1}a_{2}\ldots a_{N}$ and $b_{1}b_{2}\ldots b_{N}$
in $\mathcal{C}$, the $N$ concatenations $a_{2}a_{3}\ldots a_{N}b_{1},a_{3}a_{4}\ldots b_{1}b_{2},\ldots,a_{N}a_{1}\ldots b_{N-2}b_{N-1}$
are not in $\mathcal{C}$~\cite{guibas1978maximal}. Comma free codes enable efficient synchronization protocols, as one is able to determine the starting positions of codewords without ambiguity. A major drawback of comma free codes is the need to implement an exhaustive search procedure over sequence sets to decide whether or not a given string of length $n$ should be used as a codeword or not. This difficulty can be
overcome by using a special family of comma free codes, introduced
by Gilbert~\cite{gilbert1960synchronization} under the name \emph{prefix-synchronized codes}. Prefix-synchronized codes have the property that every codeword starts
with a prefix $P=p_{1}p_{2}\ldots p_{n}$, which is followed by a constrained sequence $c_{1}c_{2}\ldots c_{s}$. Moreover, for any codeword $p_{1}p_{2}\ldots p_{n}c_{1}c_{2}\ldots c_{s}$
of length $n+s$, the prefix $P$ \emph{does not appear} as a substring of $p_{2}\ldots p_{n}c_{1}c_{2}\ldots c_{s}p_{1}p_{2}\ldots p_{n-1}.$ More precisely, the constrained sequences of prefix-synchronized codes avoid the pattern $P$ which is used as the address.

Due to the choice of mutually uncorrelated addresses at \emph{large Hamming distance}, we encode each information block by \emph{avoiding only one of the address sequences}, used for that particular block. 

To explain how to perform encoding, assume that $P=p_{1}p_{2}\ldots p_{n}\in\left\{ \mathtt{A,T,G,C}\right\} ^{n}$
is a self-uncorrelated sequence. This guarantees that $p_{1}\neq p_{n}.$ Without loss of generality, let $p_{1}=\mathtt{A}$ and $p_{n}=\mathtt{G},$
and define
\begin{align*}
\bar{P}_{i} & =\left\{ \mathtt{A,C,T}\right\} \setminus\left\{ p_{i}\right\} \\
P^{i} & =p_{1}\ldots p_{i},
\end{align*}
for all $1\leq i\leq n$. In addition, assume that the elements of $\bar{P}_{i}$ are arranged 
in increasing order, say using the lexicographical ordering $\mathtt{A\prec C\prec T}$. We subsequently
use $\bar{p}_{i,j}$ to denote the $j$-th smallest element in $\bar{P}_{i}$,
for $1\leq j\leq\left|\bar{P}_{i}\right|.$ For example, if $\bar{P}_{i}=\left\{ \mathtt{C,T}\right\},$ 
then $\bar{p}_{i,1}=\mathtt{C}$ and $\bar{p}_{i,2}=\mathtt{T}.$ 

Next, we define a sequence of integers $G_{n,1},G_{n,2},\ldots$ that
satisfies the following recursive formula
\[
G_{n,\ell}=\begin{cases}
3^{\ell}, & 1\leq \ell<n,\\
\sum_{i=1}^{n-1}\left|\bar{P}_{i}\right|G_{n,\ell-i}, & \ell \geq n.
\end{cases}
\]
For an integer $\ell \geq0$ and $y<3^{\ell}$, let $\theta_{\ell}\left(y\right)=\left\{ \mathtt{A,T,C}\right\} ^{\ell}$
be a length-$\ell$ ternary representation of $y$. Conversely, for each
$W\in\left\{ \mathtt{A,T,C}\right\} ^{\ell}$, let $\theta^{-1}\left(W\right)$
be the integer $y$ such that $\theta_{\ell}\left(y\right)=W.$ 
Every integer $0 \leq x<G_{n,\ell}$ can be mapped into a sequence
of $n+\ell$ symbols $\left\{ \mathtt{A,T,C,G}\right\} $ via an encoding
algorithm that consists of two parts: $\mathtt{EncodePSC}(P,\ell,x)$ and
$\mathtt{CodePSC}(P,\ell,x)$. Algorithm $\mathtt{EncodePSC}(P,\ell,x)$ calls $\mathtt{CodePSC}(P,\ell,x)$
and returns the concatenation of $P$ and $\mathtt{CodePSC}(P,\ell,x)$. 

The steps of the encoding procedure are 
listed in Algorithm~1, where $C_{\ell}^{P}=\left\{ \mathtt{EncodePSC}(P,\ell,x)\mid0\leq x<G_{n,\ell}\right\}$, and where $n$ denotes the length of the sequence $P$. The decoding steps are described in the same chart.
\begin{table*}[ht]
\small
\begin{tabular*}{\hsize}
{@{\extracolsep{\fill}}rrrrrrr}
\hline
\multicolumn{2}{l}{\textbf{Algorithm 1} Prefix-synchronized encoding and decoding}\cr
\hline
\multicolumn{1}{l}{$X=\mathtt{EncodePSC}\left(P,\ell,x\right)$}\cr
\multicolumn{1}{l}{$\quad$ return $P\mathtt{CodePSC}\left(P,\ell,x\right);$}\cr
\hline
\multicolumn{1}{l}{$X=\mathtt{CodePSC}\left(P,\ell,x\right)$}&
\multicolumn{4}{l}{$x=\mathtt{DecodePSC}\left(P,X\right)$}\cr
\multicolumn{1}{l}{begin}&
\multicolumn{4}{l}{begin}\cr
\multicolumn{1}{l}{1$\quad$$n=\textrm{length}\left(P\right);$}&
\multicolumn{4}{l}{1$\quad$$n=\textrm{length}\left(P\right);$}\cr
\multicolumn{1}{l}{2$\quad$ if $\left(\ell \geq n\right)$}&
\multicolumn{4}{l}{2$\quad$$\ell=\textrm{length }\left(X\right);$}\cr
\multicolumn{1}{l}{3$\quad\quad$$t:=1;$}&
\multicolumn{4}{l}{3$\quad$$X=X_{1}X_{2}\ldots X_{\ell};$}\cr
\multicolumn{1}{l}{4$\quad\quad$$y:=x;$}&
\multicolumn{4}{l}{4$\quad$if $\left(\ell<n\right)$}\cr
\multicolumn{1}{l}{5$\quad\quad$ while $\left(y\geq\left|\bar{P}_{t}\right|G_{n,\ell-t}\right)$}&
\multicolumn{4}{l}{5$\quad\quad$ return $\theta^{-1}\left(X\right);$}\cr
\multicolumn{1}{l}{6$\quad\quad\quad$$y:=y-\left|\bar{P}_{t}\right|G_{n,\ell-t};$}&
\multicolumn{4}{l}{6$\quad$else}\cr
\multicolumn{1}{l}{7$\quad\quad\quad$$t++;$}&
\multicolumn{4}{l}{7$\quad\quad$find$\left(s,t\textrm{ such that }P^{t-1}\bar{p}_{t,s}=X_{1}\ldots X_{t}\right);$}\cr
\multicolumn{1}{l}{8$\quad\quad$ end;}&
\multicolumn{4}{l}{8$\quad\quad$return $\left(\sum_{i=1}^{t-1}\left|\bar{P}_{i}\right|G_{n,\ell-i}\right)+\left(s-1\right)G_{n,\ell-t}+\mathtt{DecodePSC}\left(P,X_{t+1}\ldots X_{\ell}\right);$}\cr
\multicolumn{1}{l}{9$\quad\quad\enspace$$a:=\left\lfloor \frac{y}{G_{n,\ell-t}}\right\rfloor$;}&
\multicolumn{4}{l}{9$\quad$end;}\cr
\multicolumn{1}{l}{10$\quad\quad$$b:=\textrm{mod }\left(y,G_{n,\ell-t}\right)$;}&
\multicolumn{4}{l}{end;}\cr
\multicolumn{1}{l}{11$\quad\quad$return $P^{t-1}\bar{p}_{t,a+1}\mathtt{CodePSC}\left(P,\ell-t,b\right);$}\cr
\multicolumn{1}{l}{12$\quad$ else}\cr
\multicolumn{1}{l}{13$\quad\quad$ return $\mathcal{\theta}_{\ell}\left(y\right);$}\cr
\multicolumn{1}{l}{14$\quad$ end;}\cr
\multicolumn{1}{l}{end;}\cr
\hline
\end{tabular*}
\label{table:alg}
\end{table*}

The following theorems are proved in the Supplementary Information.

\begin{thm} \label{thm:encoding}
$C_{\ell}^{P}$ is a prefix-synchronized codeword. \end{thm}


\begin{thm} \label{thm:decoding}
\label{thm:decodable}The algorithm \textup{$\mathtt{EncodePSC}(P,\ell,x)$}
outputs a uniquely decodable string, for any $0\leq x<G_{n,\ell}.$\end{thm}

A simple example describing the encoding and decoding procedure for the short address string 
$P=\mathtt{AGCTG}$, which can easily be verified to be self-uncorrelated, is provided in the Supplementary Information.

The previously described $\mathtt{EncodePSC}(P,\ell,x)$ algorithm imposes
no limitations on the length of a prefix used for encoding. This feature
may lead to unwanted cross hybridization between address primers used
for selection and the prefixes of addresses encoding the information. One approach
to mitigate this problem is to \textquotedblleft{}perturb\textquotedblright{} long prefixes in the encoded information 
in a controlled manner. For small-scale random access/rewriting experiments, the recommended approach is to
first select all prefixes of length greater than some predefined threshold.
Afterwards, the first and last quarter of the bases of these long prefixes
are used unchanged while the central portion of the prefix string
is cyclically shifted by half of its length.
For example, for the address primer $\mathtt{ACTAACTGTGCGACTGATGC}$,
the suffix $\mathtt{ACTAACTGTGCGACTG}$ produced by $\mathtt{EncodePSC}(P,\ell,x)$
maps to $\mathtt{ACTAATGCCTGGACTG}$. The process of shifting applied to this string is illustrated
below:
\[
\begin{array}{c}
\mathtt{ACTAA}\underbrace{\mathtt{CTGTGC}}\mathtt{GACTG}\\
\overset{\textrm{cyclically shift by 3}}{\Downarrow}\\
\mathtt{ACTAA}\overbrace{\mathtt{TGCCTG}}\mathtt{GACTG}
\end{array}
\]

For an arbitrary choice of the addresses, this scheme may not allow for 
unique decoding $\mathtt{EncodePSC}(P,\ell,x)$. However, there exist simple conditions 
that can be checked to eliminate primers that do not allow this transform to be ``unique''. 
Given the address primers created for our random access/rewriting experiments, we were able to uniquely map 
each modified prefix to its original prefix and therefore uniquely decode the readouts.

As a final remark, we would like to point out that prefix-synchronized coding also supports error-detection and 
limited error-correction. Error-correction is achieved by checking if each substring of the sequence represents a 
prefix or ``shifted'' prefix of the given address sequence and making proper changes when needed. 

\vspace{-0.2in}

\section{Discussion} \label{sec:discussion}
We described a new DNA based storage architecture that enables accurate random access and cost-efficient rewriting. The key component of our implementation is a new collection of coding schemes and the adaptation of random-access enabling codes from classical storage systems. 
In particular, we encoded information within blocks with unique addresses that are prohibited to appear anywhere else in the encoded information, thereby removing any undesirable cross-hybridization problems during the process of selection and amplification. We also performed four access and rewriting experiments without readout errors, as confirmed by post-selection and rewriting Sanger sequencing. The current drawback of our scheme is high cost, as synthesizing long DNA blocks is expensive. Cost considerations also limited the scope of our experiments and the size of the prototype, as we aimed to stay within a budget comparable to that used for other existing architectures. Nevertheless, the benefits of random access and other unique features of the proposed system compensate for this high cost, which we predict will decrease rapidly in the very near future. 
\section{acknowledgments}
This work was partially supported by the Strategic Research Initiative of University of Illinois,
Urbana-Champaign, and the NSF STC on Science of Information, Purdue University. A provisional patent for rewritable,
random-access DNA-based storage was filed with the University of Illinois in November 2014.

\beginsupplement


\section*{Supplementary Information}
\setcounter{section}{0}
\subsection*{List of sections}
\begin{enumerate}
\item Encoding Wikipedia Entries \textendash{} A Working Example (Section~\ref{sec:working}).
\item Proofs of Theorems (Section~\ref{sec:proofs}).
\item Address Sequences (Section~\ref{sec:addresses}).
\item Example of Encoding and Decoding Procedure (Section~\ref{sec:encdec}).
\item Experimental Synthesis, Access and Rewrite of DNA Storage Sequences (Section~\ref{sec:experiments}).
\item Hybrid DNA-Based and Classical Storage (Section~\ref{sec:hybrid}).
\end{enumerate}

\section{Encoding Wikipedia entries: A Working Example} \label{sec:working}

In this section we describe the data format used for encoding two files of size $17$ KB containing 
the introductory sections of Wikipedia pages of six universities: Berkeley, Harvard, MIT, Princeton, Stanford, and University of Illinois Urbana-Champaign. There were $1,933$ words in the text, out of which $842$ were distinct. Note that in our context, words are elements of the text separated by a space. For example, \textquotedblleft{}university\textquotedblright{} and
\textquotedblleft{}university.\textquotedblright{} are counted as two different words, 
while \textquotedblleft{}Urbana-Champaign\textquotedblright{} is counted as a single word. 
These $1,933$ words were mapped to $\left\lceil \frac{1933}{72}\right\rceil =27$
DNA blocks of length $1000$ bps, as we grouped six words into fragments, and combined $12$ fragments for prefix-synchronized encoding. 
Table \ref{tab:1_comp} provides the word counts in the files and encoding lengths (in bits) of the of the outlined procedure. 

Assume that instead of using a prefix-synchronized code, we used classical 
ASCII encoding without compression to encode the same Wikipedia pages. 
The total number of characters in the text equals $12,874$, and each character is 
mapped to a binary string of length $7$. Hence, one would need $12874\times7=90118$ bits to represent the 
data, which is equivalent to $\left\lceil \frac{90118}{2\times960}\right\rceil =47$ DNA blocks
of length $1000$ bps if we set aside two unique address flags for
the blocks. As one can see, prefix-synchronized codes offer an almost $1.7$-fold improvement in 
description length compared to ASCII encoding. This comes at the cost of storing a larger dictionary, as one encodes words rather than symbols of the alphabet. For the working example, one would require roughly $70$-times larger dictionaries, as there are $1933$ words with an average of $5.1$ symbols per word. This increased in the dictionary is not a significant problem, as only one copy of the dictionary is ever needed.

\begin{center}
\begin{table}
\centering{}%
\begin{tabular}{|>{\centering}p{3cm}|>{\centering}p{2.5cm}|>{\centering}p{2.5cm}|>{\centering}p{3cm}|>{\centering}p{3cm}|}
\cline{2-5} 
\multicolumn{1}{>{\centering}p{3cm}|}{} & \# symbols & \# distinct symbols & \# bits/distinct symbol & \# bits\tabularnewline
\hline 
Characters & $12874$ & $51$ & $6$ & $77244$\tabularnewline
\hline 
Words & $1933$ & $842$ & $12$ & $23196$\tabularnewline
\hline 
\end{tabular}\caption{\label{tab:1_comp} Comparison between character and word based encoding. Note the the number of bits per distinct symbol for the word encoding case is computed as the ceiling of the logarithm of the number of distinct symbols plus one, where the extra bit is used to prevent very small integers from being used in prefix-synchronized coding. Such integers may produce long runs of the first symbol in the address, which should be avoided. Furthermore, to ensure fixed length encoding, and hence avoid catastrophic error propagation, we doubled the number of bits used for encoding to $24$.}
\end{table}
\par\end{center}


\section{Proofs of Theorems} \label{sec:proofs}

\emph{Proof of Theorem \ref{thm:bounds}}. The proof consists of two parts. First, we prove the upper bound on $u\left(n\right)$ in Lemma 1,
and then proceed to prove a lower bound in Lemma 2. Recall that $u(n)$ denotes the largest possible size for a set of mutually uncorrelated
words of length $n$.

\textbf{Lemma 1.} \label{lem:upper_bound}
\emph{Let $u(n)$ the largest set of distinct mutually uncorrelated sequences of length $n$. Then
\[
u\left(n\right)\leq 9 \cdot 4^{n-2}.
\]}

\emph{Proof:} To prove the lemma, let us introduce some terminology. Let $d_H(\cdot,\cdot)$ stand for the Hamming distance between two words, and define the Hamming ball of radius $d$ around a point
$W$ in $\left\{\mathtt{A, T, G, C}\right\}^n$ as
\[B\left(W,d\right)=\left\{W^{\prime} \in \left\{\mathtt{A, T, G, C}\right\}^n \, : \, d_H\left(W,W^{\prime}\right)\leq d\right\}.\]
Furthermore, let
\[C\left(W,d\right)=\left\{ W^{\prime}  \in  \left\{\mathtt{A, T, G, C}\right\}^n \, : \, W^{\prime}\in B\left(W,d\right),\textrm{ \ensuremath{W^{\prime}}, \ensuremath{W}\ are correlated}\right\}\]
denote the set of sequences correlated with $W$ that are also at most at Hamming distance $d$ from $W$. 

We claim that for $n\geq d+2\geq4$, one has 
\begin{equation}\label{eq:correlation_lower_bound}
\left|C\left(W,d\right)\right|\geq2\sum_{i=0}^{d-1}\left(\begin{array}{c}
n-1\\
i
\end{array}\right)3^{i}-\sum_{i=0}^{d-2}\left(\begin{array}{c}
n-2\\
i
\end{array}\right)3^{i}.
\end{equation}

To prove the result, assume without loss of generality that $W$
starts with the symbol $\mathtt{A}$, i.e., $W=\mathtt{A}W_{2}\ldots W_{n}$.
Next, consider two scenarios regarding the structure of 
$W=\mathtt{A}W_{2}\ldots W_{n}$:
\begin{itemize}
\item $W_{n}\neq\mathtt{A}:$ In this case, any word $W^{\prime}$
in $B\left(W,d\right)$ that starts with $W_{n}$ or ends
with $\mathtt{A}$ is an element of $C\left(W,d\right).$

Let $S=\left\{ W^{\prime}:W^{\prime}\in B\left(W,d\right),\textrm{ \ensuremath{W^{\prime}} starts with }W_{n}\right\} $
and $E=\left\{ W^{\prime}:W^{\prime}\in B\left(W,d\right),\textrm{ \ensuremath{W^{\prime}}ends with }\mathtt{A}\right\} .$

Clearly, $\left|S\right|=\left|E\right|=\sum_{i=0}^{d-1}\left(\begin{array}{c}
n-1\\
i
\end{array}\right)3^{i}$ and $\left|S\cap E\right|=\sum_{i=0}^{d-2}\left(\begin{array}{c}
n-2\\
i
\end{array}\right)3^{i}.$ Therefore, $\left|C\left(W,d\right)\right|\geq\left|S\cup E\right|=2\sum_{i=0}^{d-1}\left(\begin{array}{c}
n-1\\
i
\end{array}\right)3^{i}-\sum_{i=0}^{d-2}\left(\begin{array}{c}
n-2\\
i
\end{array}\right)3^{i}.$
\item $W_{n}=\mathtt{A}:$ In this case, any word $W^{\prime}$
in $B\left(W,d\right)$ which starts or ends with $\mathtt{A}$ is also an element of 
$C\left(W,d\right)$. Using an argument similar to the one described for the previous scenario, one can show that 
$\left|C\left(W,d\right)\right|\geq2\sum_{i=0}^{d}\left(\begin{array}{c}
n-1\\
i
\end{array}\right)3^{i}-\sum_{i=0}^{d}\left(\begin{array}{c}
n-2\\
i
\end{array}\right)3^{i}.$ 

Moreover, it is straightforward to see that 
\[2\sum_{i=0}^{d}\left(\begin{array}{c}
n-1\\
i
\end{array}\right)3^{i}-\sum_{i=0}^{d}\left(\begin{array}{c}
n-2\\
i
\end{array}\right)3^{i}>2\sum_{i=0}^{d-1}\left(\begin{array}{c}
n-1\\
i
\end{array}\right)3^{i}-\sum_{i=0}^{d-2}\left(\begin{array}{c}
n-2\\
i
\end{array}\right)3^{i}.\]
\end{itemize}

For any mutually uncorrelated set $\left\{ X_{1},\ldots,X_{m}\right\} $ of size $m$, 
we have $X_{i}\notin C\left(X_{1},n\right)$, for $2\leq i\leq m.$
This implies that 
\[ \left\{ X_{1},\ldots,X_{m}\right\} \subseteq\left\{ \mathtt{A,T,C,G}\right\} ^{n}\setminus C\left(X_{1},n\right).\]

At the same time, the previous claim suggests that
\begin{align*}
\left|C\left(X_{1},n\right)\right| & \geq2\sum_{i=0}^{n-1}\left(\begin{array}{c}
n-1\\
i
\end{array}\right)3^{i}-\sum_{i=0}^{n-2}\left(\begin{array}{c}
n-2\\
i
\end{array}\right)3^{i}\\
 & =2\cdot4^{n-1}-4^{n-2}.
\end{align*}
Therefore, $m\leq4^{n}-\left(2\cdot4^{n-1}-4^{n-2}\right)=9\cdot4^{n-2}$, which completes the proof.

\textbf{Lemma 2.} \emph{Let $u(n)$ the largest set of distinct mutually uncorrelated sequences of length $n$. Then
\[
u\left(n\right)\geq4\cdot3^{\frac{n}{4}}.
\]
}

\emph{Proof:} For simplicity, assume that $m$ is even. Given a mutually uncorrelated set $\left\{ X_{1},\ldots,X_{m}\right\},$ with words of length $n$ and over the alphabet $\left\{\mathtt{A, T, G, C}\right\}$, partition $\left\{ X_{1},\ldots,X_{m}\right\} $ into two arbitrary sets $A$ and $B$ of equal size, say $A=\left\{ X_{1},\ldots,X_{\frac{m}{2}}\right\}$
and $B=\left\{ X_{\frac{m}{2}+1},\ldots,X_{m}\right\}.$ We argue that $C=\left\{ XY\mid X\in A,\,Y\in B\right\} $ is a mutually uncorrelated
set with words of length $2n$. 
\begin{itemize}
\item First, we show that the elements in $C$ are self-uncorrelated: For
an arbitrary element $Z\in C$, we have $Z=XY.$
Since the two sequences $\left\{ X,Y\right\}$ are mutually uncorrelated, 
one can easily verify that $Z_{1}^{i}\neq Z_{2n-i+1}^{2n},$
for $i\in\left\{ 1,\ldots,2n-1\right\} \setminus\left\{ n\right\}.$
Moreover, since $X\neq Y$, it holds that $Z_{1}^{n}\neq Z_{n+1}^{2n}.$
This establishes the claim.
\item Next, we argue that any two distinct elements in $C$ are uncorrelated:
For any two distinct elements $Z=XY$
and $Z^{\prime}=X^{\prime}Y^{\prime}$
in $C$, one can show that $Z_{1}^{i}\neq\left(Z^{\prime}\right)_{2n-i+1}^{2n}$,
for $i\in\left\{ 1,\ldots,2n-1\right\} \setminus\left\{ n\right\}$.
In addition, $X\neq Y^{\prime}$
implies that $Z_{1}^{n}\neq\left(Z^{\prime}\right)_{n+1}^{2n}.$ This completes the proof.
\end{itemize}

As a result, given a mutually uncorrelated set $\left\{ X_{1},\ldots,X_{m}\right\}$,
where $X_{i}\in\left\{ \mathtt{A,T,C,G}\right\} ^{n},$ one can construct
another mutually uncorrelated set $\left\{ Z_{1},\ldots,Z_{\frac{m^{2}}{4}}\right\}$,
where $Z_{i}\in\left\{ \mathtt{A,T,C,G}\right\} ^{2n}$. Therefore,
$u\left(2n\right)\geq\frac{u^{2}\left(n\right)}{4}.$ Observing that for $n=4$ it is possible to 
construct the following set of $12$ mutually uncorrelated sequences 
\begin{align*}
\left\{ \mathtt{ATGC,ATAC,GTAC,GTGC}\right.\\
\mathtt{ATTC,GTTC,AGGC,AAAC}\\
\mathtt{\left.GAAC,GGGC,ATTT,GTTT\right\} }
\end{align*}
establishes the base of a recursive procedure which gives $u\left(n\right)>4\cdot\left(1.31\right)^{n}.$ Note that
this bound is constructive, and the concatenation procedure preserves normalized minimum Hamming distances.

We now turn our attention to prefix-synchronized coding, and describe a number of results relevant for our subsequent discussion.

\begin{thm} [\cite{guibas1978maximal}] Given a positive
integer $N$, chose the unique integer $n=n\left(N\right)$ so that
$\beta=N2^{-n}$ satisfies
\[
\log2\leq\beta<2\log2.
\]
Then, the maximal prefix-synchronized code of length $N$ has cardinality
\[
N^{-1}2^{N-1}\beta e^{-\beta}\left(1+o\left(1\right)\right),\textrm{ as }N\rightarrow\infty,
\]
for a prefix of the form $10\ldots0$.
\end{thm}

Note that the above results indicate that codes avoiding one address sequence represent an exponentially large family of
binary sequences. We prove a similar result for the case of $4$-ary sequences that avoid a set of $m$ mutually uncorrelated 
sequences. To establish the claim, we need the following definitions. Let $g\left(0\right),g\left(1\right),\ldots,$ be an integer sequence over a finite alphabet. 
Define the generating function of the sequence
\begin{equation}
G\left(z\right)=\sum_{N=0}^{\infty}g\left(N\right)z^{-N}.\label{eq:generating function} \notag
\end{equation}
\begin{thm} \label{thm:generating}
Suppose that $\left\{ X_{1},\ldots,X_{m}\right\} $ is a set
of mutually uncorrelated sequences of length $n$ over the alphabet $\left\{ \mathtt{A,T,C,G}\right\}$. 
Let $f\left(N\right)$, with $f\left(0\right)=1$, be the number of strings of length $N$ over
$\left\{ \mathtt{A,T,C,G}\right\} $ that do not contain substrings in $\left\{ X_{1},\ldots,X_{m}\right\}$. Then
\begin{equation}
F\left(z\right)=\frac{z^{N}}{m+\left(z-4\right)z^{N-1}},\label{eq:size_code} \notag
\end{equation}
where $F\left(z\right)$ is the generating function of the sequence $\{{f\left(N\right)\}}$. 
\end{thm}

\emph{Proof of Theorem \ref{thm:generating}}. The result is a direct consequence of Theorem 4.1 of~\cite{guibas1978maximal}. For $1\leq i\leq m,$ let
$f_{i}\left(n\right)$ denote the number of strings of length $n$
over $\left\{ \mathtt{A,T,C,G}\right\} $ that contain no element of $\left\{ X_{1},\ldots,X_{m}\right\}$, 
except for a single copy of $X_{i}$ at the right-hand side of the string. 
Let $F_{i}\left(z\right)$ be the generating function of $f_{i}\left(n\right)$. 
Then, we have the following system of equations that holds for the two sets of aforementioned functions:
\begin{align}
\left(z-4\right)F\left(z\right)+zF_{1}\left(z\right)+\ldots+zF_{m}\left(z\right)=z\nonumber \\
F\left(z\right)-z\left(X_{1}\circ X_{1}\right)_{z}F_{1}\left(z\right)-z\left(X_{2}\circ X_{1}\right)_{z}F_{2}\left(z\right)-\ldots-z\left(X_{m}\circ X_{1}\right)_{z}F_{m}\left(z\right)=0\nonumber \\
\vdots\nonumber \\
F\left(z\right)-z\left(X_{1}\circ X_{m}\right)_{z}F_{1}\left(z\right)-z\left(X_{2}\circ X_{m}\right)_{z}F_{2}\left(z\right)-\ldots-z\left(X_{m}\circ X_{m}\right)_{z}F_{m}\left(z\right)=0\label{eq:1}
\end{align}
By using the fact that $\left(X_{i}\circ X_{i}\right)_{z}=z^{n-1}$,
for $1\leq i\leq m$, and $\left(X_{i}\circ X_{j}\right)_{z}=0$,
for $1\leq i\neq j\leq m$, one can show that
\begin{equation}
F\left(z\right)=z^{n}F_{1}\left(z\right)=\ldots=z^{n}F_{m}\left(z\right).\label{eq:2}
\end{equation}
The result follows by replacing (\ref{eq:1}) into the
first line of (\ref{eq:2}). 

As the dominant pole of the generating function is close to $4$, the number of sequences avoiding a set of mutually uncorrelated sequences grows roughly as $4^n$.

\emph{Proof of Theorem \ref{thm:encoding}}. Since $P$ is self-uncorrelated, we need to show that this string
is not contained in the output of $\mathtt{CodePSC}(P,\ell,x)$, where
the output of $\mathtt{CodePSC}(P,\ell,x)$ equals
\[
\mathtt{CodePSC}(P,\ell,x)=P^{t_{1}-1}\bar{p}_{t_{1},s_{1}}\ldots P^{t_{r}-1}\bar{p}_{t_{r},s_{r}}\theta_{t_{0}}\left(\cdot\right),
\]
for some input $\theta_{t_{0}}\left(\cdot\right),$ and $1\leq t_{0},t_{1},\ldots,t_{r}<n$.
Consequently, if $P$ is a substring of the output of $\mathtt{CodePSC}(P,\ell,x)$,
then the last symbol of $P$ (recall that we assumed this symbol to be $\mathtt{G}$)
has to appear in one of the following three positions:
\begin{itemize}
\item The symbol appears in $P^{t_{i}-1},$ for a unique $1\leq i\leq r:$ In this case, 
there exists a suffix of $P$ appearing as a prefix of $P^{t_{i}-1}.$
This contradicts our assumption that $P$ is self uncorrelated. 
\item The symbol appears in $\bar{p}_{t_{i},s_{i}},$ for a unique $1\leq i\leq r:$ 
This contradicts our assumption that $\bar{p}_{t_{i},s_{i}}\neq\mathtt{G}.$
\item The symbol appears in $\theta_{t_{0}}\left(\cdot\right):$ This contradicts
our assumption that $\mathtt{G}$ does not appear in
$\theta_{t_{0}}\left(\cdot\right)\in\left\{ \mathtt{A,T,C}\right\} ^{t_{0}}$. 
\end{itemize}
Therefore, the string $P$ does not appear as a substring in the output
of $\mathtt{CodePSC}(P,m,x)$, which completes the proof.

\emph{Proof of Theorem \ref{thm:decoding}}. It suffices to show that the output of $\mathtt{CodePSC}(P,\ell,x)$
is uniquely decodable. We use induction arguments to establish this
result. For the basis step, by the definition of the output of $\mathtt{CodePSC}$,
it is straightforward to show that $\mathtt{CodePSC}(P,\ell,x)$ returns
the encoding $\theta_{\ell}\left(x\right)$, which represents a one-to-one
mapping from $0\leq x<3^{\ell}$ to $\left\{ \mathtt{A,T,C}\right\} ^{\ell}$
whenever $\ell<n$. For the inductive step, we assume that the result
is true for all $\ell<r$, as well as for all $r \geq n$, and show that
it is consequently true for $\ell=r$. 

For $\ell=r$, $\mathtt{CodePSC}(P,\ell,x)$ returns
\[
P^{t-1}\bar{p}_{t,s}\mathtt{CodePSC}(P,\ell-t,b),
\]
for some integer values $s,b$ and for some $1\leq t<n,$ where $x=\left(\sum_{i=1}^{t-1}\left|\bar{P}_{i}\right|G_{n,\ell-i}\right)+\left(s-1\right)G_{n,\ell-t}+b.$
Therefore $x$ is uniquely decodable if and only if $s,t$ and $b$
are unique. Since sequences of the form $P^{t-1}\bar{p}_{t,s}$ are
prefix-free one can uniquely identify both $t$ and $s.$ Moreover
$\ell-t<r$, hence by the induction hypothesis it follows that $b$ is
also uniquely decodable from $\mathtt{CodePSC}(P,\ell-t,b)$. Hence,
$x$ can be uniquely decoded.  

\begin{center}
\begin{table}
\centering{}%
\begin{tabular}{|>{\centering}p{3cm}|>{\centering}p{8cm}|}
\hline 
Desgination of primer & Sequence \tabularnewline
\hline 
B1-forward & $5'\mathtt{AATTACTAAGCGACCTTCTC}3'$\tabularnewline
\hline 
B1-reverse & $5'\mathtt{ACTTATTGCGACTTCTAAGG}3'$\tabularnewline
\hline 
gBlock-B1-reverse & $5'\mathtt{CTTCATAACAACTAACTGTGAC}3'$\tabularnewline
\hline 
B1-SU1-reverse & $\begin{array}{c}
5'\mathtt{CGTGCACTCATAACCCATATTTCAAGAGCT}\\
\mathtt{AGCTATTCCTCTCCCTTAAAAGTAAATGAC}3'
\end{array}$\tabularnewline
\hline 
B1-SD1-forward & $\begin{array}{c}
5'\mathtt{GGGAGAGGAATAGCTAGCTCTTGAAATAT}\\
\mathtt{GGGTTATGAGTGCACGATCATCACATAAC}3'
\end{array}$\tabularnewline
\hline 
B2-forward & $5'\mathtt{AACCTAACCATCTTCCTCTC}3'$\tabularnewline
\hline 
B2-reverse & $5'\mathtt{AAACGATCCCCTGACAGAGC}3'$\tabularnewline
\hline 
gBlock-B2-forward & $5'\mathtt{GAAGCACAGTGTTGCTGCGTG}3'$\tabularnewline
\hline 
B2-SU1-reverse & $\begin{array}{c}
5'\mathtt{CAGCTTGTATCCCATCTCAACCCTAATTC}\\
\mathtt{CATAACCGTCAGCGCAGTTGACTAGTCTC}3'
\end{array}$\tabularnewline
\hline 
B2-SD1-forward & $\begin{array}{c}
5'\mathtt{CTGCGCTGACGGTTATGGAATTAGGGTT}\\
\mathtt{GAGATGGGATACAAGCTGATATGGGAAC}3'
\end{array}$\tabularnewline
\hline 
B3-forward & $5'\mathtt{ATAATAGGCCTGATGATCTC}3'$\tabularnewline
\hline 
B3-reverse & $5'\mathtt{AAGAAGAACCAGTAAGCAGC}3'$\tabularnewline
\hline 
B3-SU1-reverse & $\begin{array}{c}
5'\mathtt{AACATCTACTCACTCTCAATCTAAGCTTGA}\\
\mathtt{ACTGTGTACACACCATCGCTCTTGTACGCC}3'
\end{array}$\tabularnewline
\hline 
B3-SU2-forward & $\begin{array}{c}
5'\mathtt{GTGTACACAGTTCAAGCTTAGATTGAGAGT}\\
\mathtt{GAGTAGATGTTGATGCGAGGCGAAAGATGT}3'
\end{array}$\tabularnewline
\hline 
B3-SD2-reverse & $\begin{array}{c}
5'\mathtt{GACTTCCCCCCTATAATCCATTAATGCTAG}\\
\mathtt{ATCAAGCCGCATATACTATGTTGCAAATAC}3'
\end{array}$\tabularnewline
\hline 
B3-SD2-forward & $\begin{array}{c}
5'\mathtt{GCGGCTTGATCTAGCATTAATGGATTA}\\
\mathtt{TAGGGGGGAAGTCGCTGCTGGTACTCTG}3'
\end{array}$\tabularnewline
\hline 
\end{tabular}\caption{\label{tab:2}List of primers for rewriting (editing) the blocks B1, B2 and B3. The primers for the gBlock method are listed
separately for those used with the OE-PCR method.
In the latter case, the labels of DNA fragments SU and SD stand for sample upstream and sample downstream. In OE-PCR, we linked two DNA fragments or three DNA fragments into the final PCR products; when two fragments were linked, the first fragment was labeled UP (U), while the second fragment was labeled DOWN (D); when three fragments were combined, the second fragment was labeled MIDDLE (M).}
\end{table}
\par\end{center}

\section{Address Sequences} \label{sec:addresses}

Consider the following set of strings of length $20$,
\begin{align*}
\mathtt{ACTAACTGTGCGACTGATGC}\\
\mathtt{ACACTATCGAGCTGACACGT}\\
\mathtt{AGTCAGCAGTAGTCAGTCAG}\\
\mathtt{ACTGAGCTGAGCGTATATCG}\\
\mathtt{ACTCAGCTACGACTCACATG}
\end{align*}
with GC content equal to $50\%$, i.e., $10$ GC bases. The sequences are mutually uncorrelated
and at Hamming distance exactly $10$ from each other. The sequences
do not exhibit secondary structures at room temperature, as verified by the mfold and Vienna
packages. We used these addresses for a very small-scale, proof-of-concept random access/rewriting
experiment of a $4$ KB file. 

In the large scale random access/rewriting experiment described in Section~\ref{sec:experiments}, we used
different address sequences for the two flanking ends of the $1000$ bps blocks. 
The sequences we synthesized include:

\begin{align*}
\textrm{block 1:}\left(\mathtt{CTCTTCCAGCGAATCATTAA,ACTTATTGCGACTTCTAAGG}\right)\\
\textrm{block 2:}\left(\mathtt{CTCTCCTTCTACCAATCCAA,AAACGATCCCCTGACAGAGC}\right)\\
\textrm{block 3:}\left(\mathtt{CTCTAGTAGTCCGGATAATA,AAGAAGAACCAGTAAGCAGC}\right)\\
\textrm{block 4:}\left(\mathtt{CTCTTTCGCTGTGCACAAAA,AAATCGGAAATTCGTGTCGC}\right)\\
\textrm{block 5:}\left(\mathtt{CTCTGCTGGAAATGTGTGAA,AATTCACGGTCCGAAACACC}\right)\\
\textrm{block 6:}\left(\mathtt{CTCTGTTCCTCCTTTCTCGT,TGTAGACGATTTGATTGGCG}\right)\\
\textrm{block 7:}\left(\mathtt{CTCTAGCAACTTCCGCAAAT,ACGAGATTCATACCGGACCC}\right)\\
\textrm{block 8:}\left(\mathtt{CTCTAGCTTCCCTATCCATA,TGCAGAAGAGGAGTGTCAGC}\right)\\
\textrm{block 9:}\left(\mathtt{CTCTATAGGCTCTGGTATGT,TTTAACCCGCCCGTACAGCC}\right)\\
\textrm{block 10:}\left(\mathtt{CTCTCGCTCATCTCATGTTT,ACAGTACTTGCCCAATTCGC}\right)\\
\textrm{block 11:}\left(\mathtt{CTCTGTACTCCGCTGAATCA,TAAACATTACAAGCCCCTCG}\right)\\
\textrm{block 12:}\left(\mathtt{CTCTTCTTCCCTGACGATGT,AATACAACTTCTAACCACCC}\right)\\
\textrm{block 13:}\left(\mathtt{CTCTTGATCCTACTGAGAAA,TTAATAGTTCCCGGCAGCCC}\right)\\
\textrm{block 14:}\left(\mathtt{CTCTAGTGACGTGACAGGTA,TTAGAACGAACCAGTATAGC}\right)\\
\textrm{block 15:}\left(\mathtt{CTCTACCTAAGGCCTTTGAA,TTGACCCATGAGCCAGCACC}\right)\\
\textrm{block 16:}\left(\mathtt{CTCTACAGTAGTAAACTCGT,TGCTGAACTCTAATCTGTCC}\right)\\
\textrm{block 17:}\left(\mathtt{CTCTGGGCGGCTGTACACAA,ATACACTCATAACACCTCGG}\right)\\
\textrm{block 18:}\left(\mathtt{CTCTGCGATCACAAAAAGTT,ACAACTATACGTGTCGGACC}\right)\\
\textrm{block 19:}\left(\mathtt{CTCTTTAGCACGAGTCCTAT,TGAACCCGTCGTGCTAATCG}\right)\\
\textrm{block 20:}\left(\mathtt{CTCTAATACGCACGCCCATT,ATACGGGATACAATTAGGGC}\right)\\
\textrm{block 21:}\left(\mathtt{CTCTGAGGCGTGGATATTTT,AATACATCCCTAAAAGCCGG}\right)\\
\textrm{block 22:}\left(\mathtt{CTCTGCGTGTTCATTCCATT,TGAGGATAGGATTAGTAAGG}\right)\\
\textrm{block 23:}\left(\mathtt{CTCTAAGAATCTGACTGCAT,ATGTTAACACTGAGTAAGGG}\right)\\
\textrm{block 24:}\left(\mathtt{CTCTGATCGAACCCATGTCA,ACATGACCTACATAACGTCC}\right)\\
\textrm{block 25:}\left(\mathtt{CTCTCTGGTGGCCTAAAAAT,AACAGAGATCAGAGCAGTGG}\right)\\
\textrm{block 26:}\left(\mathtt{CTCTAGAGAAACGTTGAAGT,AACCCGTACTCACTATGCCG}\right)\\
\textrm{block 27:}\left(\mathtt{CTCTGACGTCTACACAACAT,TTTGTAGATCCCAAGCATCG}\right)
\end{align*}

The pairs of sequences were used to flank the two ends of the data blocks. Only the addresses on the 
left were used for subsequent prefix-synchronized coding.

The sequences on the left-hand side of the pairing have ``interleaved'' $\left\{ \mathtt{G,C}\right\} $ and $\left\{ \mathtt{A,T}\right\} $
bases -- for example, they all start with $\mathtt{CTCT\ldots}$. This ensures a \textquotedblleft{}GC
balancing\textquotedblright{} property for the prefixes of the addresses.

\section{Encoding and Decoding Example} \label{sec:encdec}

In this section, we illustrate the encoding and decoding procedure for the short address string 
$P=\mathtt{AGCTG}$, which can easily be verified to be self-uncorrelated.

More precisely, we explain how to compute a sequence of integers $G_{n,1},G_{n,2},\ldots,G_{n,7}$, described in the main body of the paper. As before, $n$ denotes the length of the address string, which in this case equals five. 

One has
\[
\left(G_{n,1},G_{n,2},\ldots,G_{n,7}\right)=\left(3,9,27,81,267,849,2715\right).
\]

The algorithm $\mathtt{CodePSC}(P,8,550)$ produces:
\begin{flalign*}
    &550=0\times G_{5,7}+550& \\
    &\Rightarrow\mathtt{CodePSC}(P,8,550)=\underline{\mathtt{C}}\mathtt{CodePSC}(P,7,550) & \\
    &550=0\times G_{5,6}+550&\\
    &\Rightarrow\mathtt{CodePSC}(P,7,550)=\underline{\mathtt{C}}\mathtt{CodePSC}(P,6,550) &\\ 
    &550=2\times G_{5,5}+0\times G_{5,4}+16&\\
    &\Rightarrow\mathtt{CodePSC}(P,6,550)=\underline{\mathtt{AA}}\mathtt{CodePSC}(P,4,16), &\\
    &16=0\times3^{3}+1\times3^{2}+2\times3^{1}+1\times3^{0}&\\
    &\Rightarrow\mathtt{CodePSC}(P,4,16)=\underline{\mathtt{ATCT}}, &\\
    &\Rightarrow\mathtt{CodePSC}(P,8,550)=\underline{\mathtt{CCAAATCT}} &\\
\end{flalign*}
When running $\mathtt{DecodePSC}(P,X)$ on the encoded output $X=\underline{\mathtt{CCAAATCT}}$, the following steps are
executed:

\begin{flalign*}
    &\Rightarrow\mathtt{DecodePSC}(P,\mathtt{\underline{C}CAAATCT})=0\times G_{5,7}&\\
    &+\mathtt{DecodePSC}(P,\mathtt{CAAATCT})&\\
    &\Rightarrow\mathtt{DecodePSC}(P,\mathtt{\underline{C}AAATCT})=0\times G_{5,6} & \\
    &+\mathtt{DecodePSC}(P,\mathtt{AAATCT}), & \\
    &\Rightarrow\mathtt{DecodePSC}(P,\mathtt{\underline{AA}ATCT})=2\times G_{5,5}+0\times G_{5,4}&\\
    &+\mathtt{DecodePSC}(P,\mathtt{ATCT})&\\
    &\Rightarrow\mathtt{DecodePSC}(P,\mathtt{\underline{ATCT}})=16&\\ 
    &\Rightarrow \mathtt{DecodePSC}(P,\mathtt{CCAAATCT}) =2\times G_{5,5}+16=550&
\end{flalign*}

\section{Experimental Synthesis, Access and Rewrite of DNA Sequences} \label{sec:experiments}

A total of $27$ sequences of length $1000$ bps each were designed
to encode information retrieved from the Berkeley, Harvard, MIT, Princeton,
Stanford, and UIUC Wikipedia page in 2014. Except for sequence \#4, which
was rejected due to the complexity of its secondary structure, all sequences
were synthesized by IDT (Integrated DNA Technologies). In addition, $27$ corresponding 
address primers were synthesized by the same company. The address sequences of the blocks are listed 
in Section~\ref{sec:addresses}.

As a proof of concept, we performed a number of selection and editing experiments. These include selecting 
individual blocks and rewriting one of its sections, selecting three blocks and rewriting three sections in each, 
two close to the flanking ends, and one in the middle. The edits involved information about 
the budget of the institutions at a given year of operation. Detailed information about the original sequences and 
their rewritten forms is given in the following sections.

\begin{figure}
\begin{center}
\includegraphics[scale=0.6]{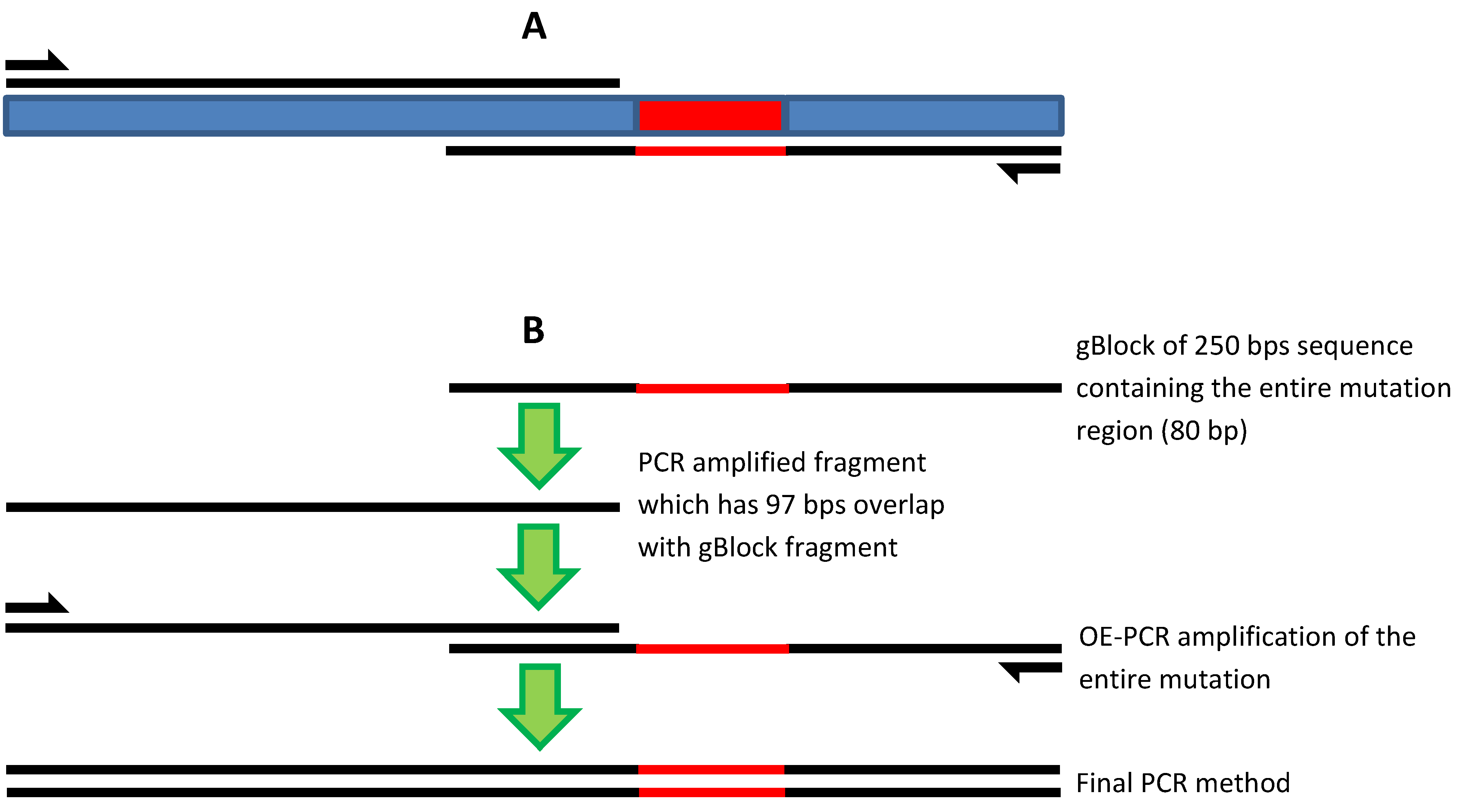}
\caption{\label{fig:2}A) Schematic depiction of the editing method using gBlocks. B) Detailed description of the generation of the mutation.
Four sequences (ranging in length from $177$ to $588$ bps) containing
the entire edit region were gBlock synthesized from IDT. The remaining
parts of the $1000$ bps sequences were PCR amplified. A homology
in at least 30 bps between the flanking end sequence of the blocks
and the corresponding end of the gBlock fragment was created. By one
OE-PCR, the desired edits were generated in a one-pot matter.}
\end{center}
\end{figure}

We denoted the blocks on which we performed selection and editing by B1, B2, and B3. The primers used for performing the edits in the blocks are listed in Table~\ref{tab:2}. Note that two primers were synthesized for each rewrite, for the forward and reverse direction. In addition, two different editing (mutation) techniques were used, gBlock and Overlap-Extension (OE) PCR; gBlocks are double-stranded genomic fragments that are frequently used as primers, for gene construction or for mediated genome editing. An illustration of editing via gBlocks is shown in Fig.~\ref{fig:2}.
On the other hand, OE-PCR is a variant of PCR used for specific DNA sequence editing via point mutations or splicing. An illustration of the procedure is given in Fig.~\ref{fig:2}. To demonstrate the plausibility of a cost efficient method for
editing, OE-PCR was used with general primers ($\leq60$ bps) only. For edits shorter than
$40$ bps, the mutation sequences were designed as overhangs in primers.
Then, the three PCR products were used as templates for the final
PCR reaction involving the entire $1000$ bps rewrite. 

All $27$ linear $1000$ bps fragments were mixed, and the mixture
was used as a template for PCR amplification and selection of the
B1, B2 and B3 sequences. The results of selection are shown in Fig \ref{fig:1},
where three banks of size $1000$ bps are depicted. These banks indicate
that sequences of the correct length were isolated. Subsequent sequencing
confirmed that the sequences were indeed the user requested B1, B2 and
B3 strands. A summary of the experiments performed is provided in Table~\ref{tab:summary}.

\begin{table}
\centering{}%
\begin{tabular}{|>{\centering}p{3cm}|>{\centering}p{2.5cm}|>{\centering}p{2.5cm}|>{\centering}p{3cm}|>{\centering}p{3cm}|}
\hline 
Sequence identifier & Number of sequence samples & Length of the edited region (in bps) & Selection accuracy / readout error percentage & Description of editing method\tabularnewline
\hline 
\hline 
B1-M-gBlock & $5$ & $20$ & $5/5/0\%$ & gBlock method\tabularnewline
\hline 
B1-M-PCR & $5$ & $20$ & $5/5/0\%$ & OE-PCR method\tabularnewline
\hline 
B2-M-gBlock & $5$ & $28$ & $5/5/0\%$ & gBlock method\tabularnewline
\hline 
B2-M-PCR & $5$ & $28$ & $5/5/0\%$ & OE-PCR method\tabularnewline
\hline 
B3-M-gBlock & $5$ & $41+29$ & $5/5/0\%$ & gBlock method\tabularnewline
\hline 
B3-M-PCR & $5$ & $41+29$ & $5/5/0\%$ & OE-PCR method\tabularnewline
\hline 
\end{tabular}\caption{\label{tab:summary} Selection, rewriting and sequencing results. Each rewritten
$1000$ bps sequence was ligated to a linearized pCRTM-Blunt vector
using the Zero Blunt PCR Cloning Kit and was transformed into \emph{E.
coli.} The \emph{E. coli} strains with correct plasmids were sequenced
at ACGT, Inc. Sequencing was performed using two universal primers:
M13F\_20 (in the reverse direction) and M13R (in the forward direction)
to ensure that the entire blocks of $1000$ bps are covered.}
\end{table}

\subsection{B1 mutation B1-M synthesis}

The unedited B1\_original (B1) sequence is of the form:
\begin{align*}
\mathtt{AATTACTAAGCGACCTTCTCGGATAGAACGCTTAGTTGGTGCGTTGACAT}\\
\mathtt{GCTCGAACTGATCATCGGTCACTTGCATTCATTATTGATTGTTGAGTTGA}\\
\mathtt{GAAGCGCATTGGTGTCACTCGTTGCTGGGTCATTTTCGGCGAGAGAAACA}\\
\texttt{GTTCACTGTGGCGTGATGTTTTGAAATGAGGGAGAGTTCTCTTAACTGCA}\\
\mathtt{GTTGGAGTTCAGTATACTCGGGATAGTGTAACAGAGGGAGGCGGATGTGT}\\
\mathtt{GTATTGATGTGAAGTCTTTCACGTGCGGGCTAGGTCGTAATGACGGGTCG}\\
\mathtt{GGAACTATTCATTGGCGCAATAGTGATTTTGATGAATGATGGATAGAACG}\\
\mathtt{CTTAAAGGGAAACTATATAGTTCAAAGCTCGTCGGCGGTGTCGAGGATGT}\\
\mathtt{ATAGGGGTTAATGAATGGTGGAACTTACTTATACTATAGATTGGACTGGT}\\
\mathtt{GGTATGAGAACTTCACTAATTATTGACGTC}\mathtt{\textcolor{red}{ACAGTTAGTTGTTATGAAGT}}\\
\mathtt{GATAATATGAATCGAGCGCAACAGGACTAGTCATTTACTTTTAAGGGAGA}\\
\mathtt{GGAATAGCTAATCTCAAATTTTTTTTATGT}\mathtt{\textcolor{red}{GAGTGCACGATCATCACATA}}\\
\mathtt{ACATAGGAGGCGATGAGACAGCGACTCAATCTGACTAATTCATTATAGGA}\\
\mathtt{GTTATATGAAGAGTTCGGAACGAAGCTAGCGCTTTCGCACAATGCGAGGG}\\
\mathtt{ATAAGAGCGGGTGCAGAGCGAAGGGTGTGAAATTGATGGTGGATAAGAAC}\\
\mathtt{TTCGCACAGTACTAGCTAGTGGGGAGAGACTTCTATGAATTCGGAGGGAT}\\
\mathtt{ACTTGATATTGATATGGGGGGATGGCGCTATTAAGCGCAGAGCGTAAGTG}\\
\mathtt{CGCTTCAAATCGAACATTGTGTAGCTAAGCAATAGAGAAATGTGGGGATT}\\
\mathtt{GAGCAGTTCGTATCGGTTCGCATGACATACTTGGGAAAATGGCAGCTTGT}\\
\mathtt{TTAAGCTAAACTGGATGAAAGGGAGGAAAAACTTATTGCGACTTCTAAGG} 
\end{align*}
where the bases written in red represent the regions we edited.

\begin{center}
\begin{figure}
\centering{}\includegraphics{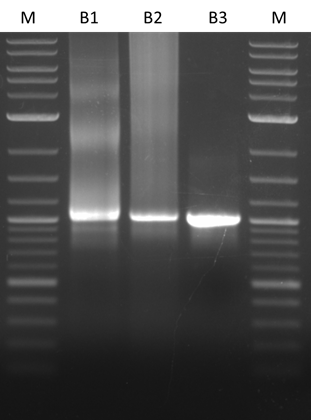}\caption{\label{fig:1}PCR of 1000 bps sequences-B1, B2, B3 from a mixture
of 26 sequences.}
\end{figure}
\par\end{center}

The edited B1\_mutation (B1\_M) sequence reads as:
\begin{align*}
\mathtt{AATTACTAAGCGACCTTCTCGGATAGAACGCTTAGTTGGTGCGTTGACAT}\\
\mathtt{GCTCGAACTGATCATCGGTCACTTGCATTCATTATTGATTGTTGAGTTGA}\\
\mathtt{GAAGCGCATTGGTGTCACTCGTTGCTGGGTCATTTTCGGCGAGAGAAACA}\\
\mathtt{GTTCACTGTGGCGTGATGTTTTGAAATGAGGGAGAGTTCTCTTAACTGCA}\\
\mathtt{GTTGGAGTTCAGTATACTCGGGATAGTGTAACAGAGGGAGGCGGATGTGT}\\
\mathtt{GTATTGATGTGAAGTCTTTCACGTGCGGGCTAGGTCGTAATGACGGGTCG}\\
\mathtt{GGAACTATTCATTGGCGCAATAGTGATTTTGATGAATGATGGATAGAACG}\\
\mathtt{CTTAAAGGGAAACTATATAGTTCAAAGCTCGTCGGCGGTGTCGAGGATGT}\\
\mathtt{ATAGGGGTTAATGAATGGTGGAACTTACTTATACTATAGATTGGACTGGT}\\
\mathtt{GGTATGAGAACTTCACTAATTATTGACGTCACAGTTAGTTGTTATGAAGT}\\
\mathtt{\textcolor{red}{GATAATATGAATCGAGCGCAACAGGACTAGTCATTTACTTTTAAGGGAGA}}\\
\mathtt{\textcolor{red}{GGAATAGCTAGCTCTTGAAATATGGGTTATGAGTGCACGATCATCACATA}}\\
\mathtt{ACATAGGAGGCGATGAGACAGCGACTCAATCTGACTAATTCATTATAGGA}\\
\mathtt{GTTATATGAAGAGTTCGGAACGAAGCTAGCGCTTTCGCACAATGCGAGGG}\\
\mathtt{ATAAGAGCGGGTGCAGAGCGAAGGGTGTGAAATTGATGGTGGATAAGAAC}\\
\mathtt{TTCGCACAGTACTAGCTAGTGGGGAGAGACTTCTATGAATTCGGAGGGAT}\\
\mathtt{ACTTGATATTGATATGGGGGGATGGCGCTATTAAGCGCAGAGCGTAAGTG}\\
\mathtt{CGCTTCAAATCGAACATTGTGTAGCTAAGCAATAGAGAAATGTGGGGATT}\\
\mathtt{GAGCAGTTCGTATCGGTTCGCATGACATACTTGGGAAAATGGCAGCTTGT}\\
\mathtt{TTAAGCTAAACTGGATGAAAGGGAGGAAAAACTTATTGCGACTTCTAAGG}
\end{align*}
with rewrites listed in red. 

\subsubsection{The gBlock method}

Since a gBlock of length longer than $500$ bps was needed, it was
more costly to synthesize the gBlock and perform rewriting than to directly re-synthesizing
the whole block. Hence, the gBlock method was not used in this case. 

\subsubsection{The OE-PCR based method}

\begin{center}
\begin{figure}
\begin{centering}
\includegraphics[scale=0.6]{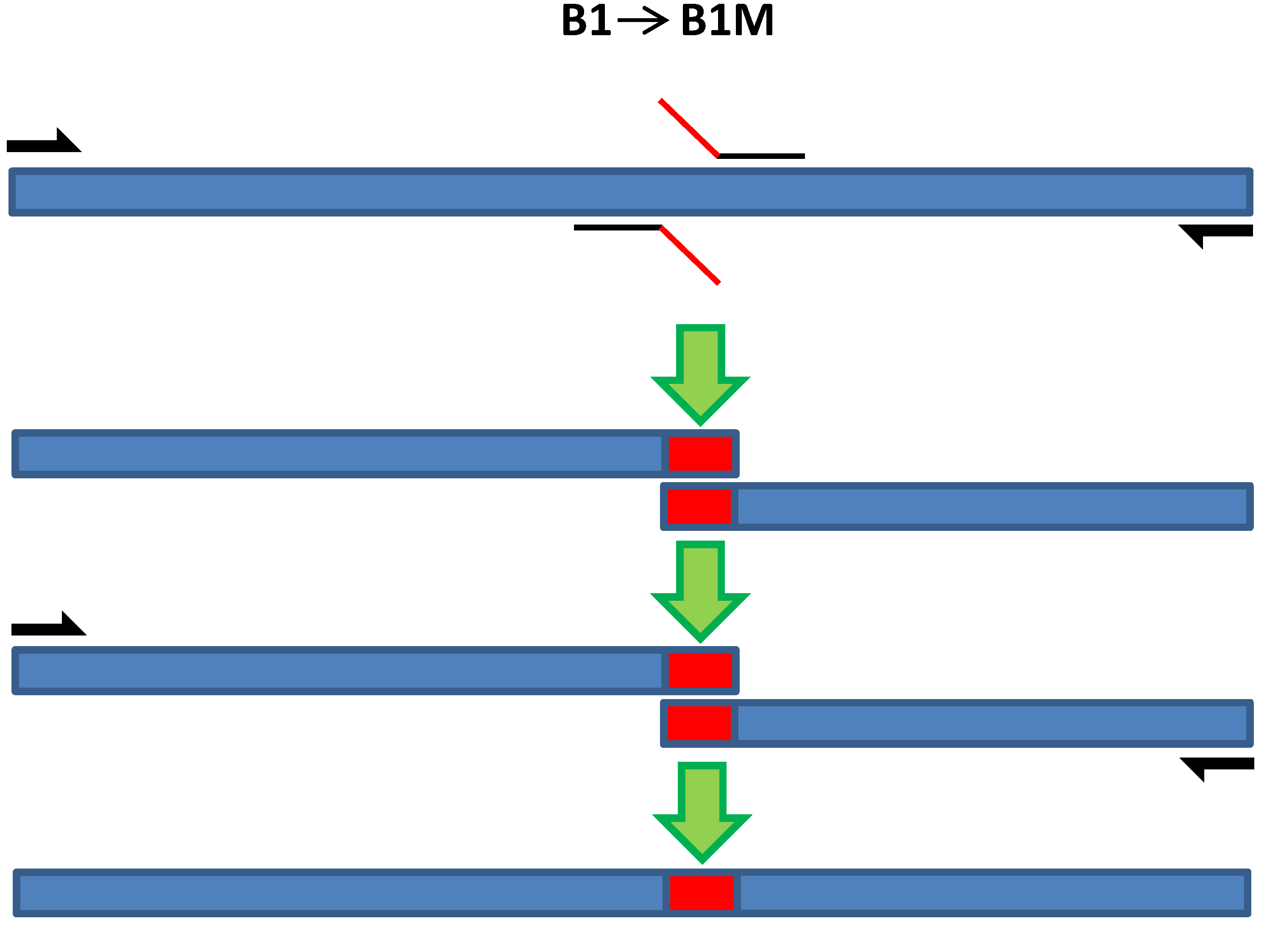}
\par\end{centering}

\caption{\label{fig:3}Illustration of the process of generating the B1 edit/mutation using general primers.}
\end{figure} 
\par\end{center}

One pair of primers was designed to PCR amplify the first portion
of the sequence B1-M. For the forward direction, the primer was 
\[
5\lyxmathsym{\textquoteright}\mathtt{AATTACTAAGCGACCTTCTC}3\lyxmathsym{\textquoteright}
\]
while for the reverse direction, the primer was 
\[
5\lyxmathsym{\textquoteright}\mathtt{CGTGCACTCATAACCCATATTTCAAGAGCTAGCTATTCCTCTCCCTTAAAAGTAAATGAC}3\lyxmathsym{\textquoteright}.
\]
The second part of the sequence was PCR amplified by using the forward
direction primer 
\[
5\lyxmathsym{\textquoteright}\mathtt{GGGAGAGGAATAGCTAGCTCTTGAAATATGGGTTATGAGTGCACGATCATCACATAAC}3\lyxmathsym{\textquoteright}
\]
and reverse direction primer
\[
5\lyxmathsym{\textquoteright}\mathtt{ACTTATTGCGACTTCTAAGG}3\lyxmathsym{\textquoteright}.
\]

Both PCR reactions used the sequence B1 as template. Two such PCR
products are shown in Fig.~\ref{fig:4}, indicating that the correct
length products were isolated in each reaction.

OE-PCR was performed in a $50$ ul reaction volume containing the
two aforementioned PCR products without primers for the first $5$
cycles and the products with primers (B1 primers in Table~\ref{tab:2})
for the later $30$ cycles. A single bank with correct size of $1000$
bps was obtained (see Fig.~\ref{fig:4}).

\subsection{B2 mutation B2-M synthesis}

The unedited B2\_original (B2) sequence is of the form:
\begin{align*}
\mathtt{AACCTAACCATCTTCCTCTCGATTTGGAGCAGATTGGTATTATTCTAGTC}\\
\mathtt{GTCGAGACTAGTCAACTGCGCTAGTTTGTGTTCATAAAATAAGAGTATGA}\\
\mathtt{GATACAAGCTGATATGGGAACTTAATTACGAAGCACAGTGTTGCTGCGTG}\\
\mathtt{GACTTGTGAAGTAGGGTGTGAGATAAGAATGATAGCGAACGCAGCGTATG}\\
\mathtt{GCTGAAGTGCTGGGCATATTGTGGTGTGGACATCTCAAAGTCTATGAAGA}\\
\mathtt{TTGGTAATAGGATGGTCTCTCGGGTCTCAAACTTCGTCAGGCAGCATTGT}\\
\mathtt{GCATGCGAGTGATTGAAAGGGAGGGTAAGGGTTATTAATAGAAAAGACTT}\\
\mathtt{ACAGGCGTTGGTATGATTCAAGATCGCAAGAATCGTGTGAGCTTGAGGAC}\\
\mathtt{TAAATAGTTTAAAGAAATAGGAATAGTTGTAATTTAAGGAGCGTGGCACG}\\
\mathtt{GATGGATCAGCGTGTCAACGGAACGCGCATTTGGGAGTTTTATGTTAAGT}\\
\mathtt{GAGCAGACTAAGGTGAAATTCAATAGTCTCTATCGTTCGAGGGTTATTGC}\\
\mathtt{TAGGGGAGACTTTGAGTGAGTGGTAATTTTGAAGCAGTATACGTAACTTT}\\
\mathtt{TTCGATTCTTAGTGGCAGTTACTCTGAATTTTAGTGTGAGCAGAGTGTGA}\\
\mathtt{TAAATAGAGAGATACGAGGTCGACACGGCTGTTGGGGGCACTTAACAGTA}\\
\mathtt{GGGGGTTGATGCTGGCGGACACTAAAGGATTTTTGAAGGGGATTGTTGGC}\\
\mathtt{GACTCACATCTAAGTGGTATTGCGGGCTCTATGAGAATCTGCTCGAGTCA}\\
\mathtt{TCTAGGTTGAGGAAGAGGGGGAGATTCTCGTTAAAGACAGTACATATTTC}\\
\mathtt{GCATACTTCTTAACGTGGAGTATGAATGTCAATGGTGGGAGATATGGGTG}\\
\mathtt{GAGGGATTTCATTCACTGCATATGTACGCTCAGGAGCGCGAACGAATCAT}\\
\mathtt{AAAACTATTGTAATATATTGATAGATAAAGAAACGATCCCCTGACAGAGC}
\end{align*}
\begin{center}
\begin{figure}
\begin{centering}
\includegraphics[scale=0.6]{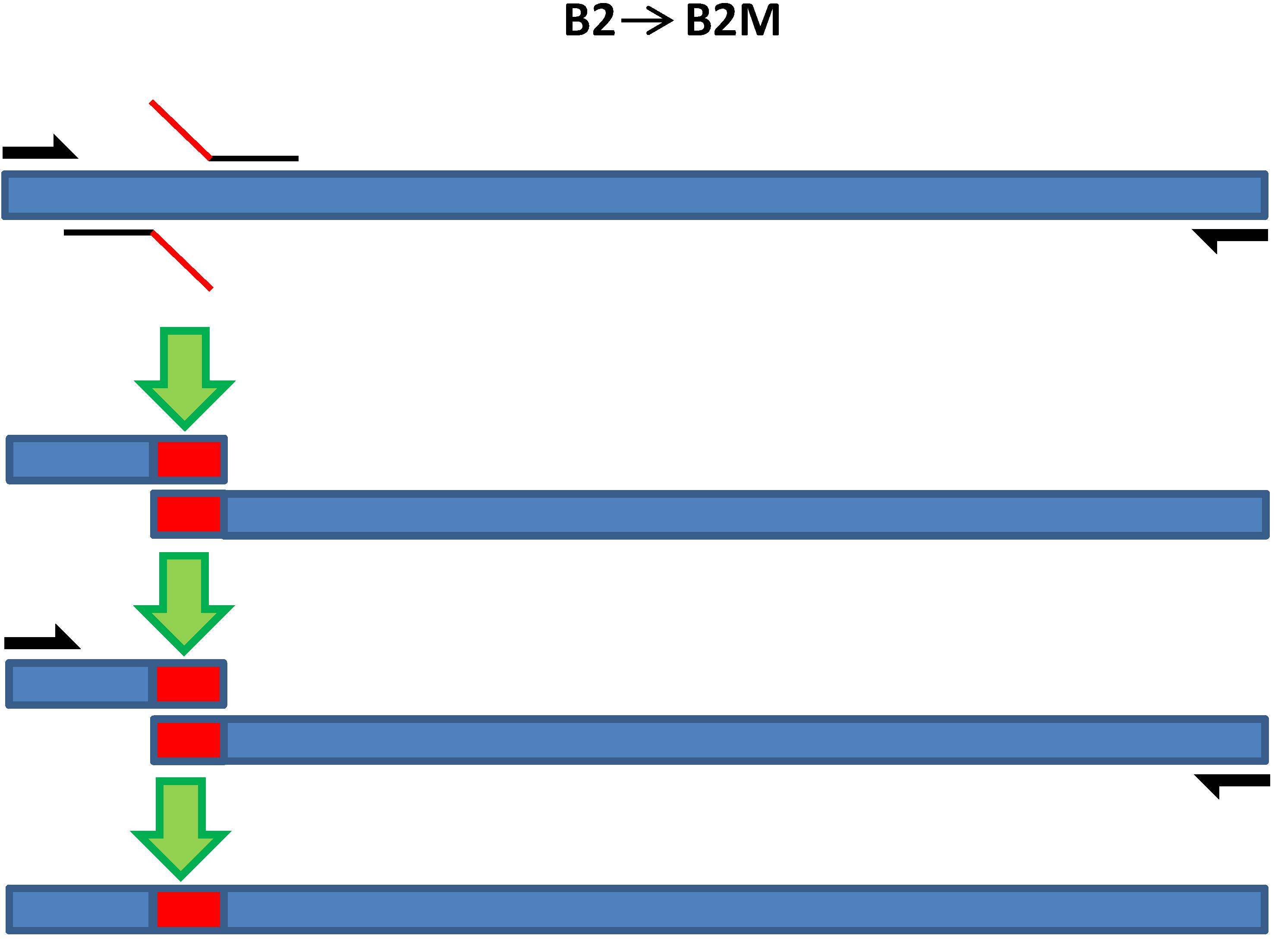}
\par\end{centering}
\caption{\label{fig:4}A schematic depiction of the process of generating the
B2 mutation using standard 60 bps primers.}
\end{figure}
\par\end{center}

The edited B2\_mutation (B2\_M) sequence is:
\begin{align*}
\mathtt{AACCTAACCATCTTCCTCTC}\mathtt{\textcolor{red}{GATTTGGAGCAGATTGGTATTATTCTAGTC}}\\
\mathtt{\textcolor{red}{GTCGAGACTAGTCAACTGCGCTGACGGTTATGGAATTAGGGTTGAGATGG}}\\
\mathtt{GATACAAGCTGATATGGGAACTTAATTACGAAGCACAGTGTTGCTGCGTG}\\
\mathtt{GACTTGTGAAGTAGGGTGTGAGATAAGAATGATAGCGAACGCAGCGTATG}\\
\mathtt{GCTGAAGTGCTGGGCATATTGTGGTGTGGACATCTCAAAGTCTATGAAGA}\\
\mathtt{TTGGTAATAGGATGGTCTCTCGGGTCTCAAACTTCGTCAGGCAGCATTGT}\\
\mathtt{GCATGCGAGTGATTGAAAGGGAGGGTAAGGGTTATTAATAGAAAAGACTT}\\
\mathtt{ACAGGCGTTGGTATGATTCAAGATCGCAAGAATCGTGTGAGCTTGAGGAC}\\
\mathtt{TAAATAGTTTAAAGAAATAGGAATAGTTGTAATTTAAGGAGCGTGGCACG}\\
\mathtt{GATGGATCAGCGTGTCAACGGAACGCGCATTTGGGAGTTTTATGTTAAGT}\\
\mathtt{GAGCAGACTAAGGTGAAATTCAATAGTCTCTATCGTTCGAGGGTTATTGC}\\
\mathtt{TAGGGGAGACTTTGAGTGAGTGGTAATTTTGAAGCAGTATACGTAACTTT}\\
\mathtt{TTCGATTCTTAGTGGCAGTTACTCTGAATTTTAGTGTGAGCAGAGTGTGA}\\
\mathtt{TAAATAGAGAGATACGAGGTCGACACGGCTGTTGGGGGCACTTAACAGTA}\\
\mathtt{GGGGGTTGATGCTGGCGGACACTAAAGGATTTTTGAAGGGGATTGTTGGC}\\
\mathtt{GACTCACATCTAAGTGGTATTGCGGGCTCTATGAGAATCTGCTCGAGTCA}\\
\mathtt{TCTAGGTTGAGGAAGAGGGGGAGATTCTCGTTAAAGACAGTACATATTTC}\\
\mathtt{GCATACTTCTTAACGTGGAGTATGAATGTCAATGGTGGGAGATATGGGTG}\\
\mathtt{GAGGGATTTCATTCACTGCATATGTACGCTCAGGAGCGCGAACGAATCAT}\\
\mathtt{AAAACTATTGTAATATATTGATAGATAAAGAAACGATCCCCTGACAGAGC}
\end{align*}
where, as before, red letters were used to indicate the rewritten region. 

\subsubsection{The gBlock method}

A $177$ bps sequence, containing the entire edited region and the
B2 string, was gBlock synthesized by IDT. Another part of B2 was PCR
amplified using the forward primer 
\[
5\lyxmathsym{\textquoteright}\mathtt{GAAGCACAGTGTTGCTGCGTG}3\lyxmathsym{\textquoteright}
\]
and reverse primer
\[
5\lyxmathsym{\textquoteright}\mathtt{AAACGATCCCCTGACAGAGC}3\lyxmathsym{\textquoteright}
\]
The B2 sequence served as a template. See Fig.~\ref{fig:4} for an illustration.

\subsubsection{The OE-PCR based method}

Over extension PCR (OE-PCR) was performed in a $50$ ul reaction volume
containing the above $177$ bps gBlock product and PCR products without
primers for the first $5$ cycles and with B2 forward and reverse
primers listed in Table \ref{tab:2} for the subsequent $30$ cycles.

The PCR product was deposited on a gel substrate and the correct $1000$ bps band
was obtained as shown in Fig.~\ref{fig:5}.

One pair of primers was designed to PCR amplify the first part of
the sequence B2-M, with forward primer 
\[
5\lyxmathsym{\textquoteright}\mathtt{AACCTAACCATCTTCCTCTC}3\lyxmathsym{\textquoteright}
\]
and reverse primer 
\[
5\lyxmathsym{\textquoteright}\mathtt{CAGCTTGTATCCCATCTCAACCCTAATTCCATAACCGTCAGCGCAGTTGACTAGTCTC}3\lyxmathsym{\textquoteright}.
\]
The second part was PCR amplified by the forward primer 
\[
5\lyxmathsym{\textquoteright}\mathtt{CTGCGCTGACGGTTATGGAATTAGGGTTGAGATGGGATACAAGCTGATATGGGAAC}3\lyxmathsym{\textquoteright}
\]
and reverse primer 
\[
5\lyxmathsym{\textquoteright}\mathtt{AAACGATCCCCTGACAGAGC}3\lyxmathsym{\textquoteright}.
\]
Both PCRs used B2 as a template. Two PCR products are shown in Fig.
\ref{fig:5}. 


\begin{center}
\begin{figure}
\begin{centering}
\includegraphics[scale=0.4]{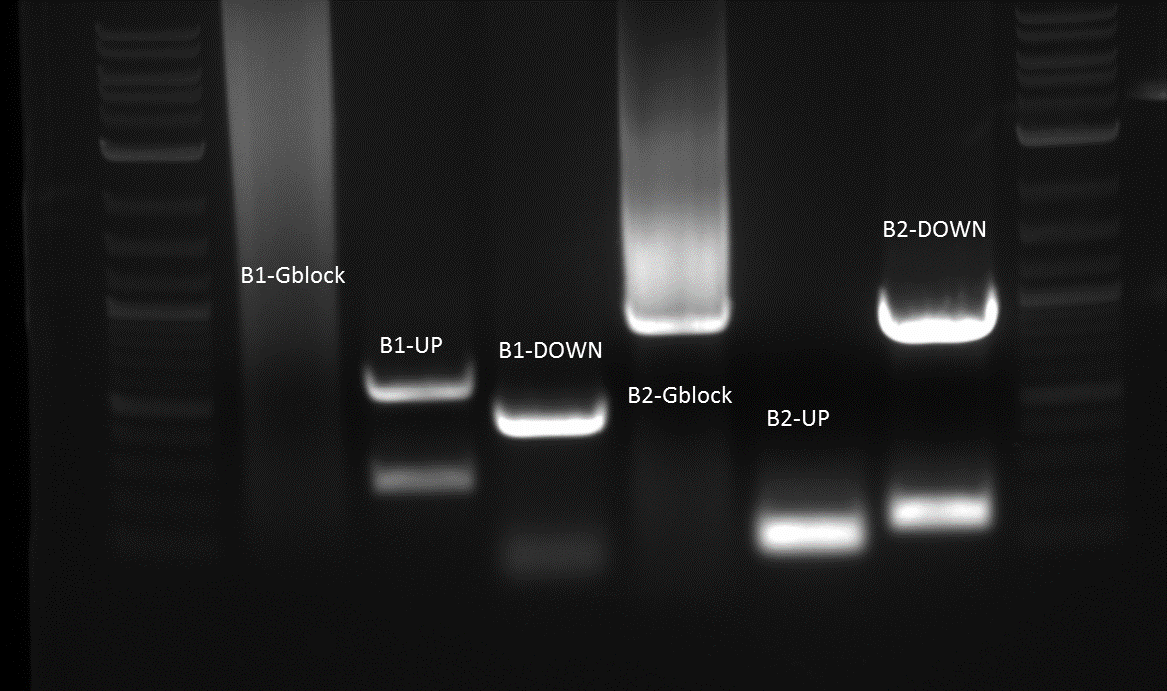}
\par\end{centering}
\caption{\label{fig:5}PCR products of B1 and B2.}
\end{figure}
\par\end{center}
\begin{center}
\begin{figure}
\begin{centering}
\includegraphics[scale=0.4]{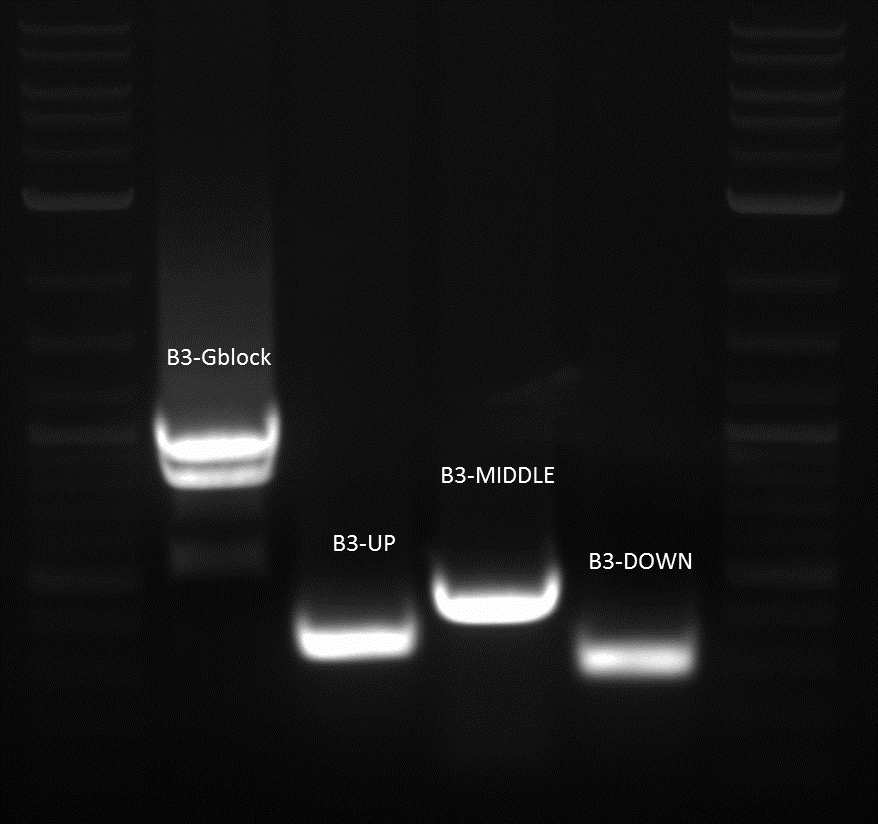}
\par\end{centering}
\caption{\label{fig:6}PCR products of B3.}
\end{figure}
\par\end{center}
\subsection{B3 mutation B3-M synthesis}

The unedited original B3 sequence equals:
\begin{align*}
\mathtt{ATAATAGGCCTGATGATCTCGATGGATGCGCGTCACTCGAGTGCGGTAGG}\\
\mathtt{CACGTCTCAGGTGATAAGTGATTGTGATTGTAGGTGAAGGGGGTAGAAAT}\\
\mathtt{GATTGAGGAAACTTGTGTACTCGTTACACGTGATAGGGTTTGATCGGCGG}\\
\mathtt{TGGAAAAATTAGGGATGGGGATAAGATTATGGGATCGTTCTCAATAATTG}\\
\mathtt{TTACGATATCGTTGTTACACAGTTGTTACGCTACGACGTCATCGATAAAG}\\
\mathtt{GTGGGTATGTGGGGGTACTATACTCTTGGGGGCGTACAAGAGCGATGGTT}\\
\mathtt{GGTCGGATTGAAATTAAAAGCATTAAGAGGTTAATTTATAGATGCGAGGC}\\
\mathtt{GAAAGATGTGAGCGCAAGTAAAGGAAACGCGAGCAAGTGATTGTTACTAA}\\
\mathtt{TTATATTAGGAGGTGATGAGGAGCGTGGTTATCTTATTGGGCGAGCTGCA}\\
\mathtt{GCGAATTCTAGATTTCTTCGAGTTACAGTCGTAGTGATGTATATAGAGTG}\\
\mathtt{GATGCGCACATTATTACATATATCGTCGAATTGGATTAGACGCAAAGAAA}\\
\mathtt{ATGCGGCATTGTAATGGGTTGTGTAAAATTGAGCGTGGTTATCTTGTCAT}\\
\mathtt{GACATAGTAAAAGTTGCTCAATTGATTGAAGCTCGATTAGGAGAAGTAAT}\\
\mathtt{TTGAAAAAAGGATAGACTAGGACTCAACGAGGAACGGGTATTTGCAACAT}\\
\mathtt{AGTATATGCGGTCTTAATCGGAGGGTAATGTTATTTGTGTGGAAGTCGCT}\\
\mathtt{GCTGGTACTCTGGGCGTTTAGGATGAATCTTCGAAACTAGGCTTTGTCAG}\\
\mathtt{AGATAGTTTGTTGGTAAGAAGAATCAGGAAACGGTAACAGAGAATAAATG}\\
\mathtt{AATTAACGTAGCAAGATTTCGTCTTTCTGGAGATGAGAAGGTGTAGTTGA}\\
\mathtt{GGAGTCGACGTTCTTTACGGAGGTGGGAGATTGGTTTTGGCAGTACTTCG}\\
\mathtt{TTAAATACACTAAAAAATTTGATAATGTAGAAGAAGAACCAGTAAGCAGC}
\end{align*}
\begin{center}
\begin{figure}
\begin{centering}
\includegraphics[scale=0.6]{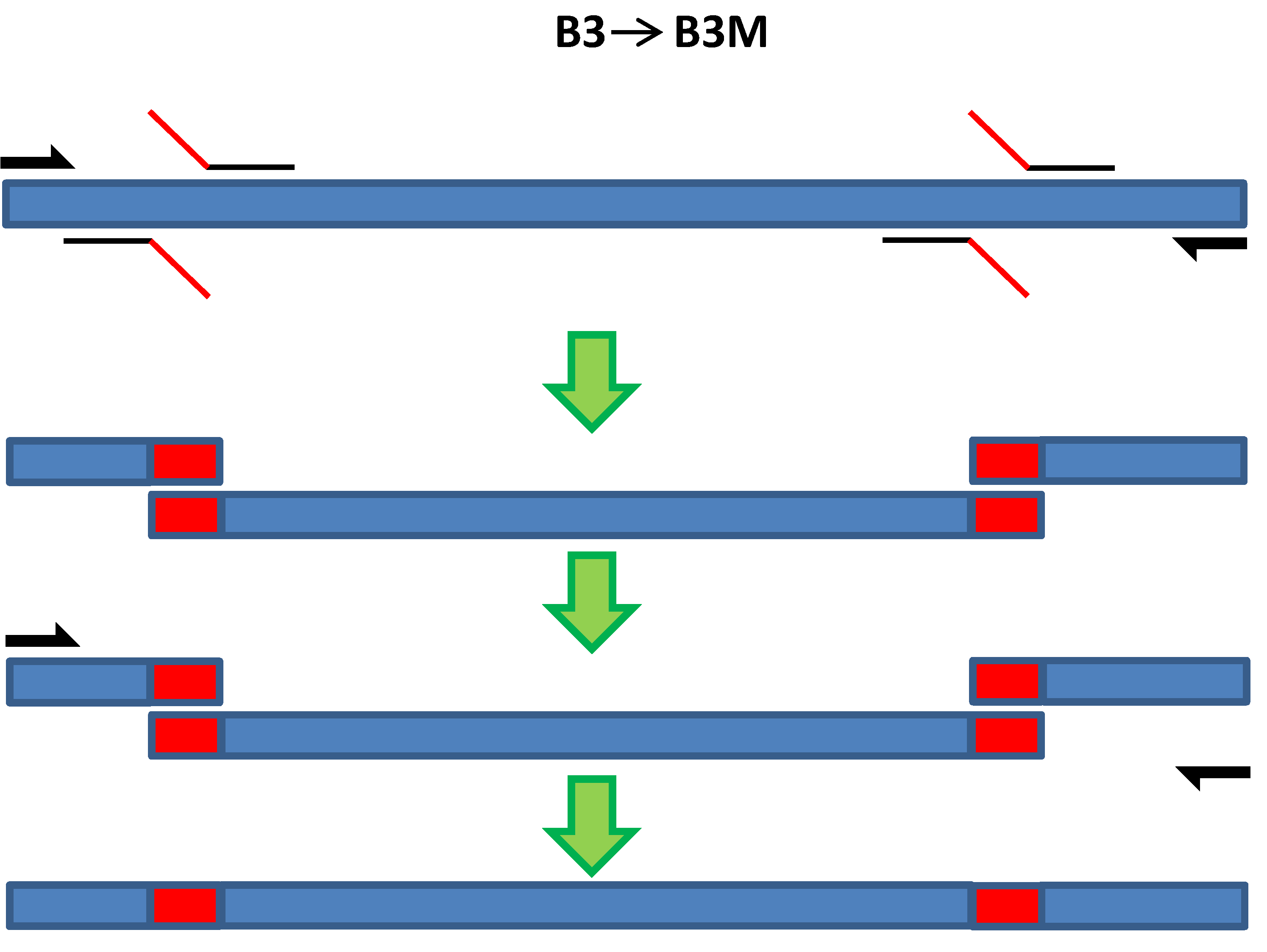}
\par\end{centering}

\caption{\label{fig:7}Scheme for generating the B3 edits using standard 60 bps
primers.}
\end{figure}

\par\end{center}
The edited sequence B3\_M mutation sequence is:
\begin{align*}
\mathtt{ATAATAGGCCTGATGATCTCGATGGATGCGCGTCACTCGAGTGCGGTAGG}\\
\mathtt{CACGTCTCAGGTGATAAGTGATTGTGATTGTAGGTGAAGGGGGTAGAAAT}\\
\mathtt{GATTGAGGAAACTTGTGTACTCGTTACACGTGATAGGGTTTGATCGGCGG}\\
\mathtt{TGGAAAAATTAGGGATGGGGATAAGATTATGGGATCGTTCTCAATAATTG}\\
\mathtt{TTACGATATCGTTGTTACACAGTTGTTACGCTACGACGTCATCGATAAAG}\\
\mathtt{GTGGGTATGT}\mathtt{\textcolor{red}{GGGGGTACTATACTCTTGGGGGCGTACAAGA}}\mathtt{\textcolor{black}{GCGATGGT}}\mathtt{\textcolor{red}{G}}\\
\mathtt{\textcolor{red}{TGTACACAGTTCAAGCTTAGATTGAGAGTGAGTAGATGTT}}\mathtt{GATGCGAGGC}\\
\mathtt{GAAAGATGTGAGCGCAAGTAAAGGAAACGCGAGCAAGTGATTGTTACTAA}\\
\mathtt{TTATATTAGGAGGTGATGAGGAGCGTGGTTATCTTATTGGGCGAGCTGCA}\\
\mathtt{GCGAATTCTAGATTTCTTCGAGTTACAGTCGTAGTGATGTATATAGAGTG}\\
\mathtt{GATGCGCACATTATTACATATATCGTCGAATTGGATTAGACGCAAAGAAA}\\
\mathtt{ATGCGGCATTGTAATGGGTTGTGTAAAATTGAGCGTGGTTATCTTGTCAT}\\
\mathtt{GACATAGTAAAAGTTGCTCAATTGATTGAAGCTCGATTAGGAGAAGTAAT}\\
\mathtt{TTGAAAAAAGG}\mathtt{\textcolor{red}{ATAGACTAGGACTCAACGAGGAACGGGTATTTGCAACAT}}\\
\mathtt{\textcolor{red}{AG}}\mathtt{TATATGCGG}\mathtt{\textcolor{red}{CTTGATCTAGCATTAATGGATTATAGGGG}}\mathtt{GGAAGTCGCT}\\
\mathtt{GCTGGTACTCTGGGCGTTTAGGATGAATCTTCGAAACTAGGCTTTGTCAG}\\
\mathtt{AGATAGTTTGTTGGTAAGAAGAATCAGGAAACGGTAACAGAGAATAAATG}\\
\mathtt{AATTAACGTAGCAAGATTTCGTCTTTCTGGAGATGAGAAGGTGTAGTTGA}\\
\mathtt{GGAGTCGACGTTCTTTACGGAGGTGGGAGATTGGTTTTGGCAGTACTTCG}\\
\mathtt{TTAAATACACTAAAAAATTTGATAATGTAGAAGAAGAACCAGTAAGCAGC}
\end{align*}
\subsubsection{The Gblock method}

Two sequences, the $560$ bps sequence containing the first mutation
region and the second $560$ bps sequence containing the second mutation
region, were gBlock synthesized by IDT. There was a $60$ bps overlap
between the two gBlocks.

\subsubsection{The OE-PCR method}

OE-PCR was performed in a $50$ ul reaction volume containing the
above two $560$ bps gBlock products without primers for the first
$5$ cycles and additional B3 forward and reverse primers listed in
Table \ref{tab:2} for the subsequent $30$ cycles. The PCR product
was deposited on a gel substrate and the correct $1000$ bps band was obtained.

One pair of primers was designed to PCR amplify the first part of
the sequence B2-M, using 
\[
5\lyxmathsym{\textquoteright}\mathtt{ATAATAGGCCTGATGATCTC3}\lyxmathsym{\textquoteright}
\]
 in the forward direction and 
\[
5\lyxmathsym{\textquoteright}\mathtt{AACATCTACTCACTCTCAATCTAAGCTTGAACTGTGTACACACCATCGCTCTTGTACGCC}3\lyxmathsym{\textquoteright}
\]
 in the reverse direction. 

The second part was PCR amplified in the forward direction by using
the primer 
\[
5\lyxmathsym{\textquoteright}\mathtt{GTGTACACAGTTCAAGCTTAGATTGAGAGTGAGTAGATGTTGATGCGAGGCGAAAGATGT}3\lyxmathsym{\textquoteright}
\]
and in the reverse direction by using the primer 
\[
5\lyxmathsym{\textquoteright}\mathtt{GACTTCCCCCCTATAATCCATTAATGCTAGATCAAGCCGCATATACTATGTTGCAAATAC}3\lyxmathsym{\textquoteright}.
\]
The third part was PCR amplified by the forward direction primer 
\[
5\lyxmathsym{\textquoteright}\mathtt{GCGGCTTGATCTAGCATTAATGGATTATAGGGGGGAAGTCGCTGCTGGTACTCTG}3\lyxmathsym{\textquoteright}
\]
and reverse direction primer 
\[
5\lyxmathsym{\textquoteright}\mathtt{AAGAAGAACCAGTAAGCAGC}3\lyxmathsym{\textquoteright}.
\]
All three PCRs used the sequence B3 as the template. All three PCR
products are shown in Fig.~\ref{fig:8}. 

OE-PCR was performed in a $50$ ul reaction volume containing the
above three PCR products without primers for the first $5$ cycles
and with B3 primers listed in Table \ref{tab:2} for the subsequent
$30$ cycles. A single bank of correct size $1000$ bps was obtained
(See Fig.~\ref{fig:9}). 

\begin{center}
\begin{figure}
\begin{centering}
\includegraphics[scale=0.4]{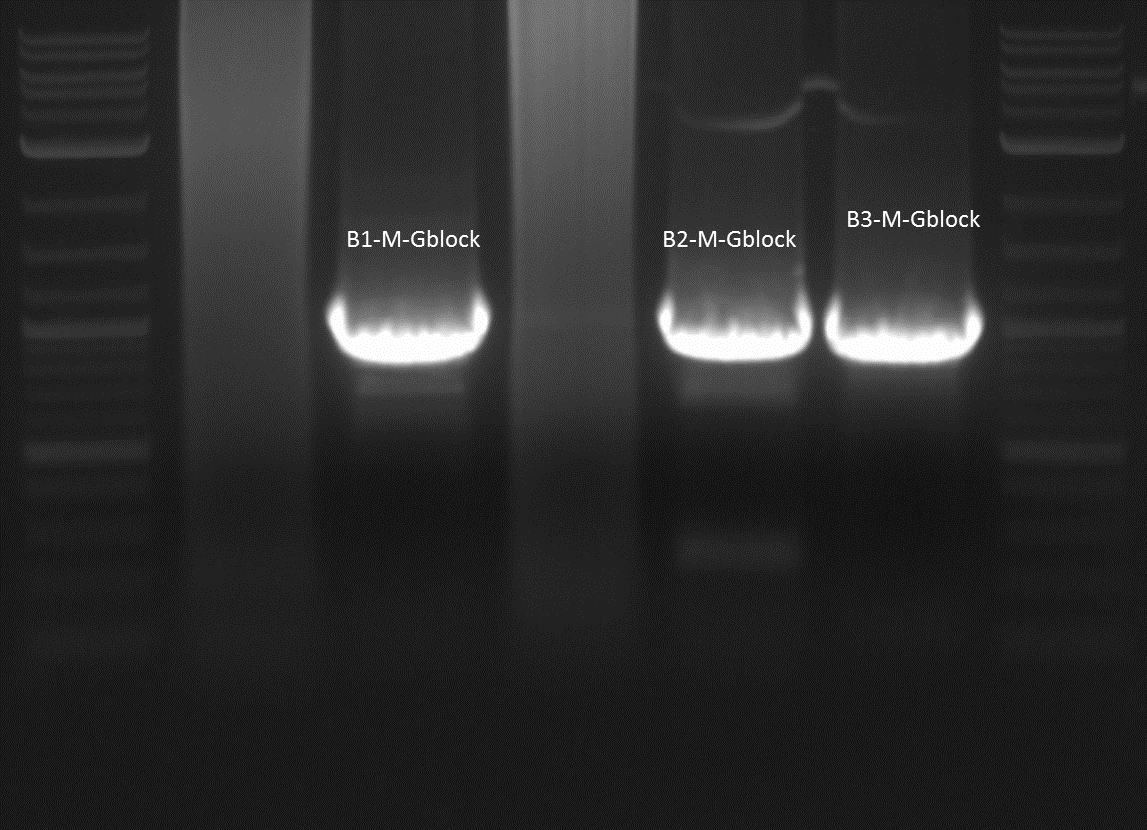}
\par\end{centering}
\caption{\label{fig:8}The generated PCR products of $1000$ bps edits from
the gBlock method, involving B1-gBlock, B2-gBlock and B3-gBlock.}
\end{figure}
\par\end{center}
\begin{center}
\begin{figure}
\begin{centering}
\includegraphics[scale=0.5]{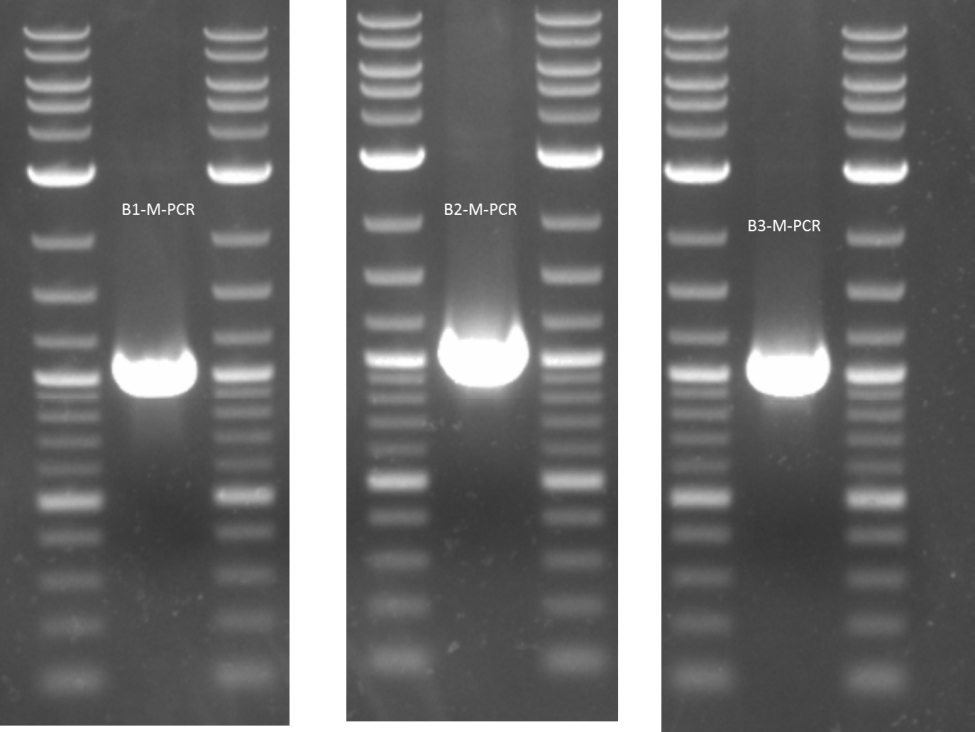}
\par\end{centering}
\caption{\label{fig:9}The generated PCR products of 1000bps sequence editing
for the OE-PCR based method, and sequences B1-PCR, B2-PCR and B3-PCR.}
\end{figure}
\par\end{center}
Correctness of the synthesized edited regions was confirmed via DNA
Sanger sequencing as follows. The PCR products of the gBlock method
and the OE-PCR method were named B1-M-gBlock, B2-M-gBlock, B3-M-gBlock
and B1-M-PCR, B2-M-PCR, B3-M-PCR, respectively. All final mutations/edits
of PCR products were purified using the QiaGen Gel Purification Kit.
The purified $1000$ bps edited sequences were blunt-ligated to the
vector named $\textrm{pCR}^{\textrm{TM}}$-Blunt (Fig. \ref{fig:10})
using the Zero Blunt PCR Cloning Kit and following the manufacturers\textquoteright{}
protocol. Five colonies of each PCR-Blunt-mutation were sent to ACTG,
Int. Sequencing was performed using two universal primers: M13F\_20
(for the reverse direction) and M13R (for the forward direction).
Bi-directional sequencing was performed in order to ensure that the
entire $1000$ bps block was completely covered. 

\begin{center}
\begin{figure}
\begin{centering}
\includegraphics[scale=0.4]{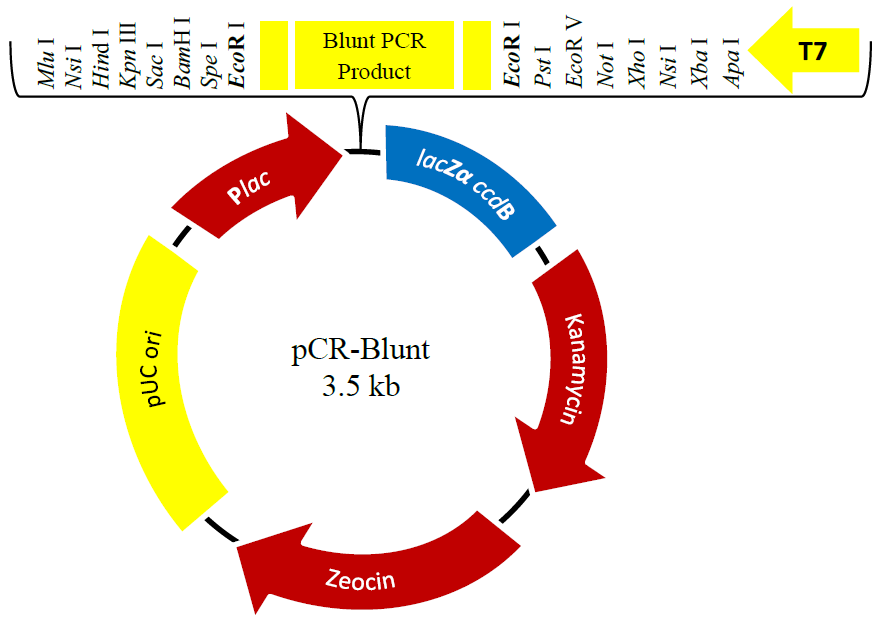}
\par\end{centering}

\caption{\label{fig:10}Map and features of PCR-Blunt vector (Life technologies).}
\end{figure}

\par\end{center}


\section{Hybrid DNA-Based and Classical Storage} \label{sec:hybrid}

In our small-scale experiments, Sanger sequencing produced two 
erroneous symbols in one strand which we were able to correct using
prefix matching. One possible problem that may arise in large scale
DNA-storage systems involving millions of blocks is erroneous sequencing which 
may not be corrected via prefix matching. In current High Throughput Sequencing technologies,
such as Illumina HiSeq or MiSeq, the dominant sources of errors are
substitutions. Due to our word grouping scheme, such substitution
errors cannot cause catastrophic error propagation, but may nevertheless accumulate
as the number of rewrite cycles increases. In this case, prefix matching
may not suffice to correct the errors and more sophisticated coding
schemes need to be used. Unfortunately, adding additional parity-check
symbols into the prefix-encoded data stream may cause problems as
the parities may violate the prefix properties and dis-balance the
GC content. Furthermore, every time rewriting is performed, the parity-checks
will need to be updated, which incurs additional cost for maintaining
the system. A simple solution to this problem is a hybrid scheme,
in which the bulk of the information is stored in DNA media, while
only parity-checks are stored on a classical device, such as flash
memory. Given that the current error-rate of short-read sequencing
technologies roughly equals $1\%$, the most suitable codes for performing
this type of coding are low-density parity-check codes~\cite{gallager1962low}.
These codes offer excellent performance in the presence of a large
number of errors and are decodable in linear time.


\end{document}